\documentclass{jfm}
\usepackage{graphicx}
\usepackage{epstopdf, epsfig}
\usepackage{pict2e}
\usepackage{tabularx}
\usepackage[percent]{overpic}
\usepackage{tikz,eso-pic} 
\usepackage{xcolor}
\usepackage{float}
\usepackage{mathtools,amssymb}
\newcommand{\abslength}{\ensuremath{L_{\mathrm{A}}}}
\newcommand{\convectivelength}{\ensuremath{L_{\mathrm{R}}}}
\newcommand{\pthree}{\ensuremath{p_2}}
\newcommand{\uvec}{\ensuremath{\mathbf{u}}}
\newcommand{\Uvec}{\ensuremath{\mathbf{U}}}
\newcommand{\uprime}{\ensuremath{\mathbf{u^{\prime}}}}
\newcommand{\pprime}{\ensuremath{p^{\prime}}}
\newcommand{\uadj}{\ensuremath{\mathbf{u^{+}}}}
\newcommand{\padj}{\ensuremath{p^{+}}}
\newcommand{\linop}{\ensuremath{\mathcal{L}}}
\newcommand{\adjop}{\ensuremath{\mathcal{L}^{+}}}
\newcommand{\freqcc}{\ensuremath{\omega_c}}
\newcommand{\dd}[1]{\mathrm{d}#1}
\newcommand{\karman}{K\'{a}rm\'{a}n}

\usetikzlibrary{shapes}

\newcommand{\NH}[1]{\textcolor{black}{#1}}

\newcommand{\JL}[1]{\textcolor{black}{#1}}

\shorttitle{Flow past cylinder rows}
\shortauthor{N. Hosseini, M. D. Griffith and J. S. Leontini}

\title{Flow states and transitions in flows past arrays of tandem
cylinders}

\author{Negar Hosseini\aff{1}
  \corresp{\email{negar.mhoseini@gmail.com}},
  Martin D. Griffith\aff{1}
 \and Justin S. Leontini\aff{1}}

\affiliation{\aff{1}Department of Mechanical and Product Design Engineering, Swinburne University of Technology,
John Street, Hawthorn VIC 3122, Australia}

\begin{document}

\maketitle

\begin{abstract}

  Direct numerical simulations at $\Rey=200$ have been conducted of
  the flow past rows of tandem cylinders.  Local stability analysis
  shows that the wakes of two tandem cylinders are characterised by
  the formation of a region where the mean flow is locally absolutely
  unstable immediately behind the second cylinder, followed by a long
  region that is convectively unstable. The location where the flow
  changes from absolutely to convectively unstable provides a
  prediction of where the placement of a third body could trigger a
  global change, which is confirmed with simulations of the flow past
  three cylinders, and the flow past two cylinders followed by a short
  flat plate.

  A third body placed downstream of the absolute/convective
  instability transition location is effectively cloaked, its presence
  having virtually no impact on the flow both upstream and
  downstream. However, when a body is placed upstream of this location
  it triggers a global change in the flow, its presence being
  broadcast throughout the flow domain.

  However, a third cylinder placed well upstream of this location does
  not trigger the global change. Sensitivity analysis of the mean flow
  is conducted, and it is shown that the third cylinder does not
  simply act as a small perturbation that can excite sensitive
  regions, but when it is very close to the second cylinder it induces
  a mean flow correction that eliminates the sensitive regions which
  may explain why it does not trigger the global change.

\end{abstract}


\section{Introduction}

Fluid flows past multiple structures which are long in a direction
perpendicular to the flow, but with a bluff or non-aerodynamic cross
section - such as cylinders - abound in engineering and nature. From
the vibration of high-voltage power lines in cross winds
\citep{tsui1986wake} and the cooling of nuclear fuel rods
\citep{lee2007}, to the mass transport through aquatic vegetation
\citep{nepf:12} and even the sensing capacity of harbor seal vibrissae
or whiskers \citep{hanke:10}, understanding the vortex formation in
such flows is vital to their control, manipulation and exploitation.

The canonical system in this class is the flow past two identical
circular cylinders. This problem is defined by two parameters, the
Reynolds number $\Rey = UD/\nu$, and the normalized distance between
the cylinder centres or pitch $p = x_c/D$, where $U$ is the free stream
velocity, $D$ is the diameter of the cylinders, $\nu$ is the kinematic
viscosity and $x_c$ is the distance between the cylinder centres.

For values of pitch $p \lesssim 3.6$, the separated shear layers from
the front cylinder reattach to the rear cylinder, and the two bodies
behave as a single, streamlined body with periodic vortex shedding in
its wake at a single distinct frequency. However for longer pitch, the
shear layers from the front cylinder roll up into distinct vortices,
and periodic vortex shedding similar to that seen from a single
isolated cylinder occurs in the gap between cylinders, and these
vortices then impinge on the rear cylinder. As a result of this
interaction, a two-row structure of vortex shedding appears behind the
second cylinder.  This process, and the critical value of pitch $p$ at
which vortex shedding in the gap occurs has been shown to be only
weakly affected by the Reynolds number over the range
$200 \leqslant \Rey < 10^5$ \citep{zdravkovich1987effects, SuPrPa00,
  hu2008flow, sumner2010two, zhou2016wake,
  griffith_lo_jacono_sheridan_leontini_2017}.

The purpose of this paper is to demonstrate two distinct phenomena
related to this two-row structure. The first is that this structure
does not share the feedback characteristics of the wake of a single
cylinder. In fact, subsequent bodies can be placed anywhere over a
large range of lengths, which we quantify below, with essentially no
impact on the flow both upstream \emph{and} downstream. Subsequent
bodies are \emph{cloaked} - their presence is not communicated to the
flow or other bodies in the array and the mean drag on these bodies
has a very small value. The second phenomena, in stark contrast to the
first, is that even small disturbances introduced closer to the second
cylinder trigger a global change in the flow, destroying the two-row
structure and completely suppressing the vortex shedding from the
first cylinder. Here, subsequent bodies are \emph{broadcast} - their
presence is signalled throughout the flow.

\JL{We analyse this behaviour using two techniques. First,} we use a
classical plane-wave, or local stability, analysis to show that the
regions of the flow where a subsequent third body's presence will be
broadcast or cloaked correlate with regions where the mean flow of the
two-cylinder system is locally absolutely or convectively unstable,
respectively. The majority of the two-row structure is reminiscent of
two free shear layers that are harmonically forced - such shear layers
produce trains of vortical structures at the same frequency as the
forcing and are known to be convectively unstable \citep{browand_1966,
  brown_1974, monkewitz_1982, ho_1984, huerre1985absolute,
  ghoniem_1987}. Our analysis shows that the majority of the two-row
structure is marginally convectively unstable, except for a small
region that is weakly absolutely unstable close to the second
body. The end of this absolutely unstable region provides a prediction
of a boundary between the broadcasting and cloaking regions. We
confirm the prediction of the stability analysis performed on the wake
of the two-cylinder system by manually placing subsequent third bodies
of varying shape in the wake. The threshold location for the most
upstream point of the third body that delineates the positions that
trigger the broadcasting or cloaking behaviour coincides closely with
the transition point from the absolutely unstable region to the
convectively unstable region. \JL{However, an upper \emph{and} lower
  boundary for the body position that can trigger a global change
  (i.e, the broadcasting pheonomenon) is found using simulation, and
  the lower boundary is not predicted with the plane wave analysis as
  the absolutely unstable region extends all the way upstream to the
  rear of the second cylinder.}

\JL{We therefore introduce an adjoint-based sensitivity analysis to
  assess the sensitivity of the time-mean wake of the two-cylinder
  system \citep{giannetti:07, luchini:14, meliga:14}. A direct
  interpretation of the sensitivity field also fails to predict the
  presence of the lower boundary. However, we present an argument that
  shows that the third cylinder cannot be treated as a small
  perturbation of the two-cylinder system, and the nonlinear
  correction of the mean flow induced by the presence of the third
  cylinder generates this lower boundary.}

\section{Methodology}

\subsection{Direct numerical simulations}
\label{sec:base}

The sharp interface immersed boundary method is used to simulate these
flows in two dimensions. The body surfaces are modelled by a set of
finite elements immersed in an underlying Cartesian grid and the
incompressible Navier-Stokes equations govern the motion of fluid:
\begin{equation}\label{eq:navier} 
  \begin{gathered}
   \frac{\partial \uvec}{\partial \tau} = - (\uvec\cdot\nabla)\uvec - \nabla P + \frac{1}{\Rey}\nabla ^2 \uvec + \mathbf{A}_b,\\ 
    \nabla\cdot\uvec=0, 
  \end{gathered}
\end{equation}

where \uvec\ is the velocity field non-dimensionalised by the
free-stream velocity $U$, $\tau = tU/D$ is time non-dimensionalised by
the advective time scale, $P$ is the pressure field
non-dimensionalised by $\rho U^2$, and $\mathbf{A}_b$ is a generic
acceleration term that models the presence of an immersed boundary. A
second-order central finite-difference scheme is used to spatially
discretise these equations. Temporal integration is performed using a
two-way time splitting scheme that results in a Poisson equation being
formed that can be solved for the pressure correction.

The flow domain in the simulation extends at least $15D$ upstream and
laterally of the centre of the array, and at least $30D$
downstream. Boundary conditions on the fluid domain are an imposed
free stream velocity and zero pressure gradient on the boundaries
upstream and transverse of the bodies, and a zero-gradient velocity
condition and zero pressure at the outlet or downstream boundary. At
the surface of the bodies, a no-slip velocity condition and zero
pressure gradient are imposed. The Cartesian grid used on the fluid
domain has 64 points across $1D$ in the vicinity of the bodies, and
the bodies are represented by 128 elements. The basic method closely
follows that presented in \citet{mittal2008versatile} and
\citet{seo2011}. Validation of the code for multiple body flows and
further details of the actual implementation used here can be found in
\citet{griffith2017sharp} and
\citet{griffith_lo_jacono_sheridan_leontini_2017}.

\subsection{Local stability analysis}

The analysis proceeds by treating the wake as a slowly-varying
parallel shear flow. A series of streamwise velocity profiles are
extracted from the mean flow at a set of downstream distances and the
linear stability of each profile is then assessed. The body of work
treating cylinder wakes in this way is reviewed in
\citet{chomaz2005global}, and there are many examples of applying this
technique to the mean flow in bluff body wakes - see
\citet{hammond1997global, pier2002frequency, thiria2007, khor2008,
  leontini2010numerical}. The relevance of the mean flow to the
dynamics of vortex wakes has also been explored in a global sense in
\citet{barkley2006linear} and \citet{arratia:14}.

Assuming viscosity plays only a secondary role in the instability and
can be neglected, the stability problem reduces to solving the
Rayleigh equation \citep{drazin2004hydrodynamic} defined as:

\begin{equation}
  \label{eqn:rayleigh}
  (\mathbf{U}_x-c)\left(\frac{\dd{^2}{\phi}}{\dd{y}{^2}}-k{^2}\phi\right)-\frac{\dd{^2}{\mathbf{U}_x}}{\dd{y}{^2}}\phi=0,
\end{equation}

where $\mathbf{U}_x$ is the streamwise velocity profile extracted from the mean
flow, $c$ is the complex wave speed, $k$ is the complex wave number,
and $\phi$ is the complex amplitude of the infinitesimal perturbation
stream function applied to the base flow. A complex frequency can be
defined as $\omega= kc$. A characteristic complex frequency \freqcc\
for each profile can be found, which is the frequency associated with
a group velocity of zero - this occurs at pinch points in the complex
$k$-plane, which coincide with cusp points in the complex
$\omega$-plane \citep{huerre1998}. The imaginary component of \freqcc\
dictates whether an instability on a particular velocity profile is
absolute or convective. If the imaginary component is positive,
disturbances grow in place and oscillate with a frequency given by the
real component of \freqcc\ - this is a local absolute instability, and
it has its own inherent dynamics with a preferred oscillation
frequency at the real component of \freqcc. If the imaginary component
is negative, disturbances do not grow in place and are instead
``washed out'' of the flow, potentially growing as they flow
downstream - this is a local convective instability, and such a region
of flow responds to any external forcing at the frequency of the
external forcing. The process of finding \freqcc\ (or equivalently the
cusp points in the complex frequency plane) is further explained in
\cite{kupfer1987} and in the context of wake flows by
\cite{leontini2010numerical}.

\subsection{\JL{Sensitivity analysis}}
\label{sec:sensitivity}

Sensitivity, or structural stability analysis, is a method for
determining which spatial regions of a flow are most sensitive to
disturbance, or structural perturbation. Such a structural
perturbation may be the introduction of some external forcing, or
simply the introduction of another object in the flow. These sensitive
regions are identified by assessing the behaviour of perturbations to
the base flow state.

Evolution equations for perturbations are formed by decomposing
the flow solution into base and perturbation components, substituting
this into the equations of motion (\ref{eq:navier}), subtracting base
flow terms and linearising the result to arrive at
\begin{equation}\label{eq:nslin} 
  \begin{gathered}
   \frac{\partial \uprime}{\partial \tau} = - \left[(\Uvec\cdot\nabla)\uprime + (\uprime\cdot\nabla)\Uvec\right] - \nabla \pprime + \frac{1}{\Rey}\nabla ^2 \uprime + \mathbf{A^{\prime}}_b,\\ 
    \nabla\cdot\uprime=0, 
  \end{gathered}
\end{equation}
where \Uvec\ is the base flow velocity field, \uprime\ and \pprime\
are the perturbation velocity and pressure fields, respectively, and
$\mathbf{A^{\prime}}_b$ is the perturbation acceleration term
introduced to account for the immersed bodies. Note the \Uvec\ is the
base flow of which the sensitivity is being assessed.

Since equation (\ref{eq:nslin}) represents a linear operator it can be
cast as
\begin{equation}
  \label{eqn:linop}
  \uprime(t+T) = \linop\uprime(t).
\end{equation}
Traditional linear stability analysis proceeds by solving for the
eigenvectors and associated eigenvalues of \linop\ such that
\begin{equation}
  \label{eqn:multiplier}
  \linop\uprime(t) = \mu\uprime(t).
\end{equation}
If any eigenvector, or mode, of \linop\ has an associated eigenvalue
or multiplier such that $|\mu| > 1$, then that mode is predicted to
grow and the base flow \Uvec\ is predicted to be unstable. These modes
are referred to as \emph{direct} or \emph{forward} modes, as they
describe the evolution of a perturbation forward in time.

The aim of the sensitivity analysis is to determine which areas of the
flow should be perturbed structurally such that the magnitude of the
multiplier associated with the leading eigenvector is reduced the most
- i.e, the areas where the direct mode and its growth rate are most
sensitive to a local perturbation. It has been well established that
this occurs in regions where the product of the local magnitude of the
direct mode and the local magnitude of the \emph{adjoint mode} is
largest \citep{giannetti:07, marquet:09, luchini:14}. The adjoint
equations associated with the linearised Navier-Stokes equations
presented in equation (\ref{eq:nslin}) can be written as
\citep{barkley:08}
\begin{equation}
  \label{eqn:adjoint}
  \begin{gathered}
    \frac{\partial \uadj}{\partial \tau} = - (\Uvec\cdot\nabla)\uadj + (\nabla\Uvec)^T\cdot\uadj - \nabla \padj + \frac{1}{\Rey}\nabla ^2 \uadj + \mathbf{A^{+}}_b,\\ 
    \nabla\cdot\uadj=0, 
  \end{gathered}
\end{equation}
where \uadj\ is the adjoint perturbation velocity field, \padj\ is the
adjoint perturbation pressure field and $\mathbf{A^{+}}_b$ is the
adjoint acceleration associated with the immersed bodies.

The adjoint equations have a sense of integrating ``backwards'' in
time, and so the equations for the adjoint can therefore be cast as
\begin{equation}
  \label{eqn:linopadj}
  \uadj(t-T) = \adjop\uadj(t).
\end{equation}
Similar to direct equations described in equation (\ref{eqn:linop}),
the eigenvectors and associated eigenvalues of \adjop\ can be
found. There will be one eigenvector of the adjoint operator \adjop\ -
an adjoint mode - associated with each eigenvector of the direct
operator \linop\ - a direct mode. The eigenvalues of the associated
adjoint and direct modes are the same.

Once the direct and adjoint modes have been resolved, the sensitivity
field $\lambda$ can be calculated by multiplying the local amplitude
of the direct and adjoint modes as
\begin{equation}
  \label{eqn:sensitivity}
  \lambda(x,y) = ||\uprime_m(x,y)|| ||\uadj_m(x,y)||
\end{equation}
where $\uprime_m$ and $\uadj_m$ are the direct and adjoint
perturbation velocity fields associated with the given mode, and the
double magnitude signifies the magnitude of a vector of complex
quantities. Note that equation (\ref{eqn:sensitivity}) is correct up
to a constant - however, the absolute magnitude of the field is not
important. The important feature to identify is the area of the flow
where $\lambda$ is large as this is where the flow is sensitive to
disturbance.

Here the focus is on identifying the the regions of the flow that,
when perturbed, can suppress the vortex shedding in the gap between
the first two cylinders. We consider the problem in the framework of a
global mode growing on the mean flow, therefore, the base flow \Uvec\
considered is the time-mean flow. We interpret suppression of the
global mode growing on this mean as the likely suppression of vortex
shedding in the gap, and search for regions of the flow where the
placement of another object - the third cylinder - may lead to this
suppression, hence the calculation of the sensitivity field of the
leading global mode on the mean flow.

Here, we have solved equations (\ref{eq:nslin}) for the direct modes,
and equations (\ref{eqn:adjoint}) for the adjoint modes using the same
spatial discretisation and time-stepping scheme as applied for solving
the Navier-Stokes equations for the base flow described in section
\ref{sec:base}. The action of the linear operators \linop\ and \adjop\
is applied by simply integrating the equations forward (or effectively
backwards for the adjoint equations) in time. The leading eigenmodes
and eigenvalues of both operators have been found using Arnoldi
iteration. For both the direct and adjoint equations, we have applied
Dirichlet boundary conditions ($\uprime = \uadj = 0$) at all the
external boundaries, effectively imposing that the perturbation decays
at long distances from any body. Similarly a no-slip condition
($\uprime = \uadj = 0$) was applied at the body surfaces. For the
pressure a Neumann boundary condition setting the normal gradient to
zero
($\partial \pprime / \partial \mathbf{n} = \partial \padj / \partial
\mathbf{n} = 0$) was applied at both the domain and body boundaries.

\subsection{\JL{Validation}}

The base Navier-Stokes solver has been validated numerous times,
including in studies of flows past pairs of cylinders in close
proximity
\citep{griffith_lo_jacono_sheridan_leontini_2017}. Similarly, the code
employed to perform the local stability analysis on extracted mean
flow profiles has also previously been employed on mean wake flow
profiles \citep{leontini2010numerical}. Therefore, the validation of
these two components is not repeated here.

The sensitivity analysis consists of three components, which are
\begin{itemize}
\item the calculation of the direct mode by timestepping equation
  (\ref{eq:nslin}), and then solving for eigenvectors and associated
  eigenvalues of equation (\ref{eqn:multiplier}) using Arnoldi
  iteration
\item the calculation of the adjoint mode by timestepping equation
  (\ref{eqn:adjoint}) \emph{backwards} in time, and then solving for
  eigenvectors and associated eigenvalues of equation
  (\ref{eqn:linopadj}) using Arnoldi iteration
\item the multiplication of the amplitude of the direct and adjoint
  modes to form the sensitivity field according to equation
  (\ref{eqn:sensitivity}).
\end{itemize}

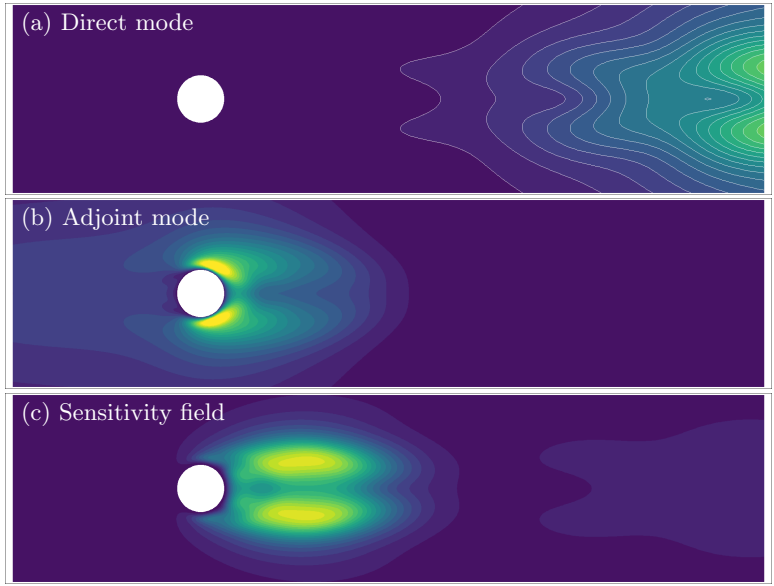
\begin{figure}\centering
  \setlength{\unitlength}{\textwidth}
  \begin{picture}(1,0.57)
    \put(0.125,0.38){\includegraphics[width=0.75\unitlength]{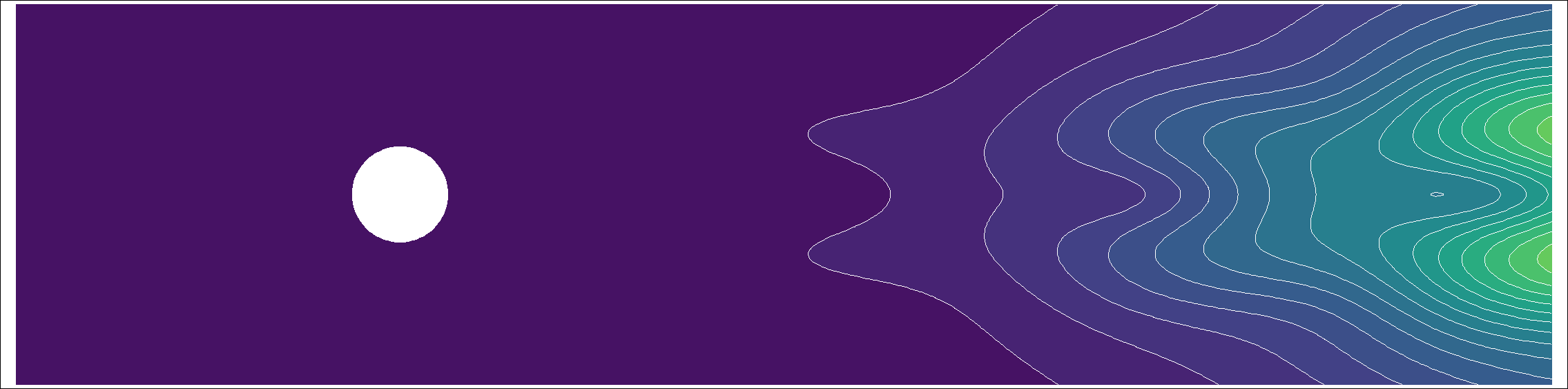}}
    \put(0.125,0.19){\includegraphics[width=0.75\unitlength]{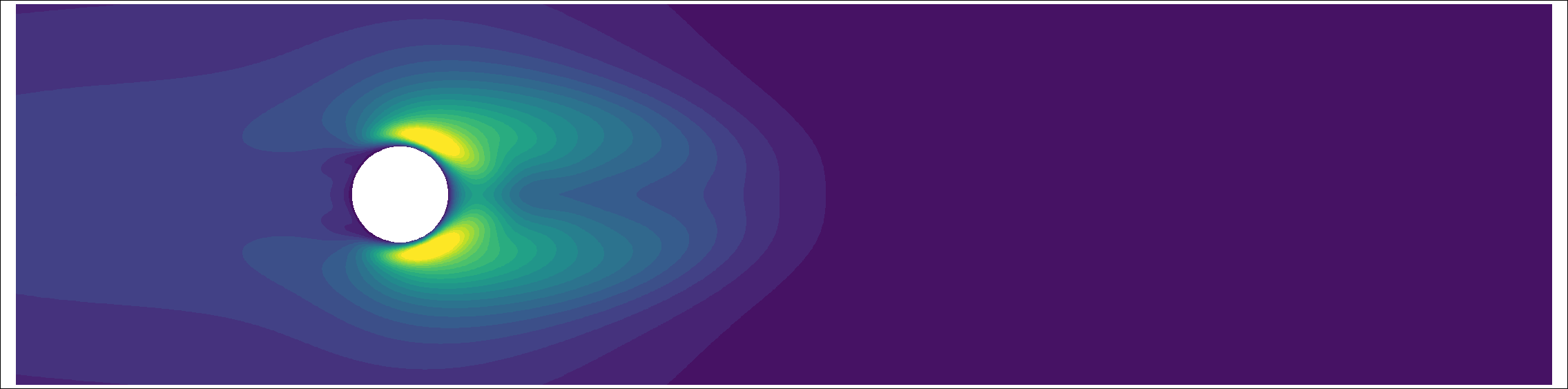}}
    \put(0.125,0.00){\includegraphics[width=0.75\unitlength]{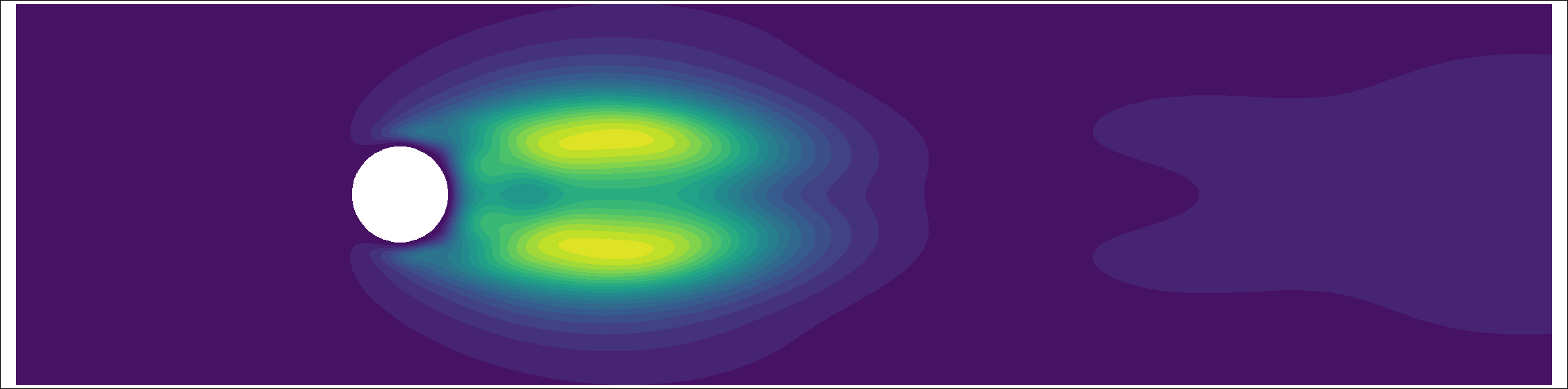}}

    \put(0.14,0.54){\textcolor{white}{(a) Direct mode}}
    \put(0.14,0.35){\textcolor{white}{(b) Adjoint mode}}
    \put(0.14,0.16){\textcolor{white}{(c) Sensitivity field}}
  \end{picture}
  \caption{\label{fig:valid_sens}Validation of the calculation of the
    sensitivity field of the leading global mode on the steady flow
    past a cylinder at $\Rey = 50$. Contours of (a) direct mode
    magnitude, (b) adjoint mode magnitude, (c) the sensitivity field
    $\lambda$. Dark/light colours represent low/high values. The
    comparison with the result of the same calculation in
    \citet{giannetti:07} is excellent.}
\end{figure}

We validate this process against the data from \citet{giannetti:07}
for the sensitivity field growing on the steady flow past a single
cylinder at $\Rey = 50$. Figure \ref{fig:valid_sens} shows contours of
the magnitude of the direct and adjoint leading modes, as well as the
final sensitivity field. The comparison of all of these fields with
the data of \citet{giannetti:07} is excellent, with the location and
relative intensity of the sensitive regions (where the sensitivity
field is high) faithfully reproduced.

We highlight that no explicit link is made between the direct and
adjoint modes - each is calculated from separate eigenvalue problems
as defined in equations (\ref{eqn:linop}) and
(\ref{eqn:linopadj}). The eigenvalue of the adjoint mode should match
that of the corresponding direct mode. Table \ref{tab:valid} shows
data related to the eigenvalues extracted for the leading modes for
the direct and adjoint problems via Arnoldi iteration. The table shows
a very close match between the eigenvalues which differ by at most
$3.5\%$ in the imaginary component, and $1.8\%$ in magnitude. The
leading modes are close to marginally stable as would be expected for
the steady flow past a cylinder at $\Rey = 50$, which is known to
become unstable around $\Rey = 47$ \citep{dusek:94}. The frequency
associated with the oscillatory modes also closely matches that
calculated by \citet{giannetti:07}. The match of the sensitivity field
structure, and the calculated eigenvalues, with previously published
data, and the self-consistency with the match between eigenvalues of
the direct and adjoint problems, provides some confidence that the
code used is reliable.

\begin{table}\centering
  \begin{tabular}{ccc}
    Quantity                  & Direct mode         & Adjoint mode        \\
    \hline
    $\mu_{T = 1}$              & $0.7438\pm 0.6631i$ & $0.7471\pm 0.6874i$ \\
    $|\mu|$                   & 0.9965              & 1.015                \\
    $f = (T/(2\pi))\tan^{-1}{(\mu_i/\mu_R)}$ &  0.1158             & 0.1183               \\
    $f_{GL}$                   & 0.1187              & 0.1187              \\
  \end{tabular}
  \caption{\label{tab:valid}Eigenvalue data for the validation
    calculation of the sensitivity field for the global mode in the
    steady flow past a cylinder at $\Rey = 50$. The multiplier
    $\mu_{T=1}$ is the ratio of the mode from one sample to the next
    where samples are separated by a period $T = 1$, generated by
    evolving the governing equations for the direct/adjoint mode
    forwards/backwards in time by and increment $T$.  The frequency
    $f$ is that associated with the oscillatory mode, and $f_{GL}$ is
    the frequency measured by \citet{giannetti:07}. }
\end{table}

\subsection{Problem set-up}

\begin{figure}
\centering
\begin{tikzpicture}

\node at (-2.5,0.75) {$  $};

\node at (-2.5,0) {$(a)$};
\node at (-2.5,-1.2) {$(b)$};
\node at (-2.5,-2.4) {$(c)$};

\node at (-0.5,0.3) {$U$};
\node at (-0.5,-.3) {$\nu$};

\node at (-0.5,-0.95) {$U$};
\node at (-0.5,-1.45) {$\nu$};

\node at (-0.5,-2.15) {$U$};
\node at (-0.5,-2.65) {$\nu$};

\draw[->] (0,0.3) -- (1.5,0.3);
\draw[->] (0,0) -- (1.5,0);
\draw[->] (0,-0.3) -- (1.5,-0.3);

\draw[->] (0,-0.9) -- (1.5,-0.9);
\draw[->] (0,-1.2) -- (1.5,-1.2);
\draw[->] (0,-1.5) -- (1.5,-1.5);

\draw[->] (0,-2.1) -- (1.5,-2.1);
\draw[->] (0,-2.4) -- (1.5,-2.4);
\draw[->] (0,-2.7) -- (1.5,-2.7);

\draw[color=black!100,  thick](3,0) circle (.35);
\draw[color=black!100,  thick](5,0) circle (.35);
\draw[color=black!100,  thick](3,-1.2) circle (.35);
\draw[color=black!100,  thick](5,-1.2) circle (.35);
\draw[color=black!100,  thick](7,-1.2) circle (.35);
\draw[color=black!100,  thick](3,-2.4) circle (.35);
\draw[color=black!100,  thick](5,-2.4) circle (.35);

\draw[color=black!100, double distance=1.5pt, line join=round, line cap=round,  thick] (6.71,-2.43)--(7.29,-2.43);

\draw[<->] (3,-.5) -- (5,-.5);            
\node at (4,-0.75) {$p$};

\draw[<->] (3,-1.7) -- (5,-1.7);
\draw[<->] (5,-1.7) -- (7,-1.7);
\node at (4,-1.95) {$p$};
\node at (6,-1.95) {$\pthree$};

\draw[<->] (3,-2.9) -- (5,-2.9);
\draw[<->] (5,-2.9) -- (7,-2.9);
\node at (4,-3.15) {$p$};
\node at (6,-3.15) {$\pthree$};

\draw[<->] (2.65,0) -- (3.35,0);
\node at (3,0.15) {$D$};
\draw[<->] (4.65,0) -- (5.35,0);
\node at (5,0.15) {$D$};
\draw[<->] (6.65,-2.28) -- (7.35,-2.28);
\node at (7,-2.1) {$D$};

\end{tikzpicture}
\caption{\label{fig:systemschematic}Schematic description of the three parts of the present study.}
\end{figure}
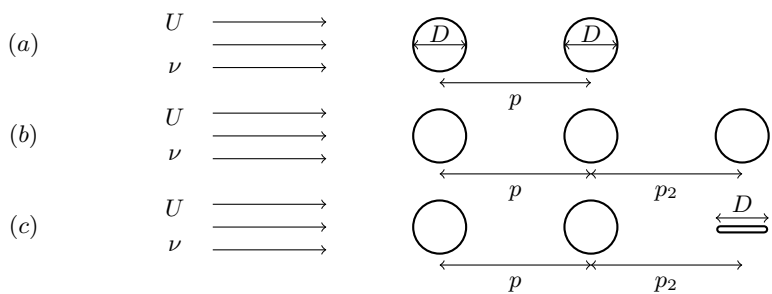

The three different configurations investigated in this study are
shown in figure~\ref{fig:systemschematic}. The cylinders and plate are
equal in size and length ($D$) and immersed in a free stream.  All the
simulations are performed for flows with constant $\Rey=200$. The
centre-to-centre distance between the bodies in the array, $p$ and
$\pthree$, are systematically varied.

\section{Results and discussion}

\subsection{Characterisation of the flows past the two-cylinder
  configuration}
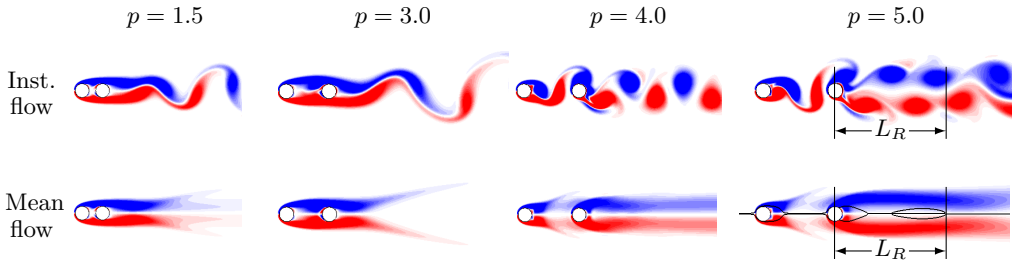
\begin{figure}
  \centering
\begin{tabular}{>{\centering\arraybackslash}m{0.7cm}m{2.5cm}m{3cm}m{3cm}m{4cm}}

    &
\hspace*{0.7cm}$p=1.5$&
\hspace*{1cm}$p=3.0$&
\hspace*{0.9cm}$p=4.0$&
\hspace*{1.1cm}$p=5.0$\\

Inst. flow&
\hspace*{-0.01cm}\includegraphics[width=.165\textwidth, trim={270 100 470 80},clip]{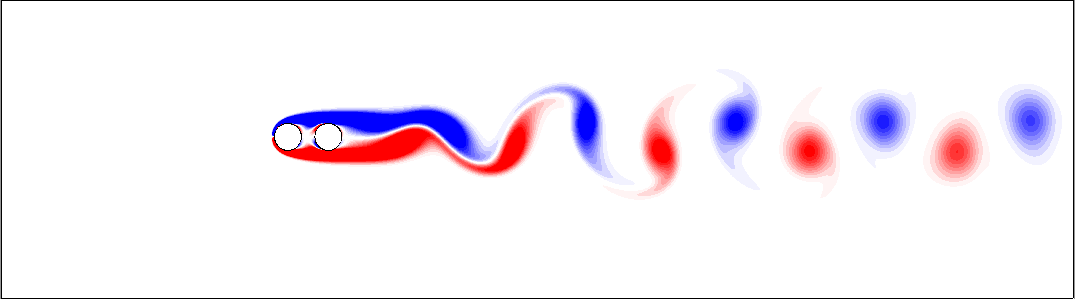}&
\hspace*{-0.01cm}\includegraphics[width=.225\textwidth, trim={19 1 188 1},clip]{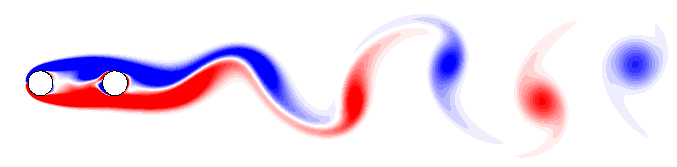}&
\hspace*{-0.1cm}\includegraphics[width=.203\textwidth, trim={17 19 230 20},clip]{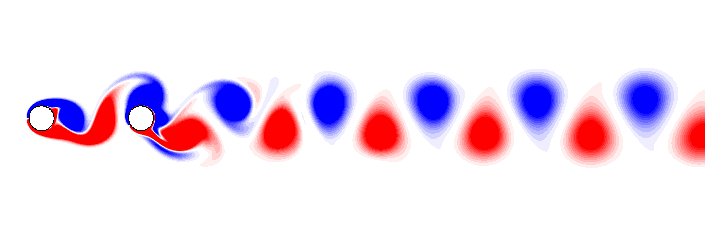}&
\hspace*{-0.18cm}\includegraphics[width=.255\textwidth, trim={20 16 200 20},clip]{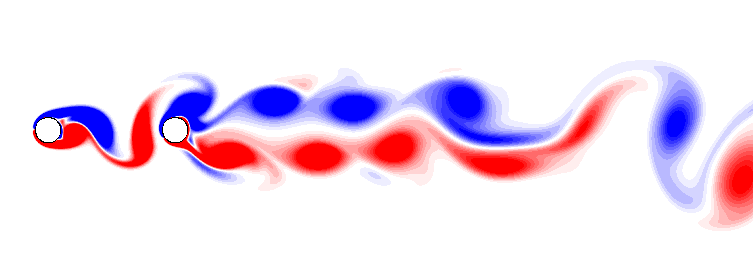}\\

Mean flow&
\hspace*{-0.01cm}\includegraphics[width=.165\textwidth, trim={270 100 470 80},clip]{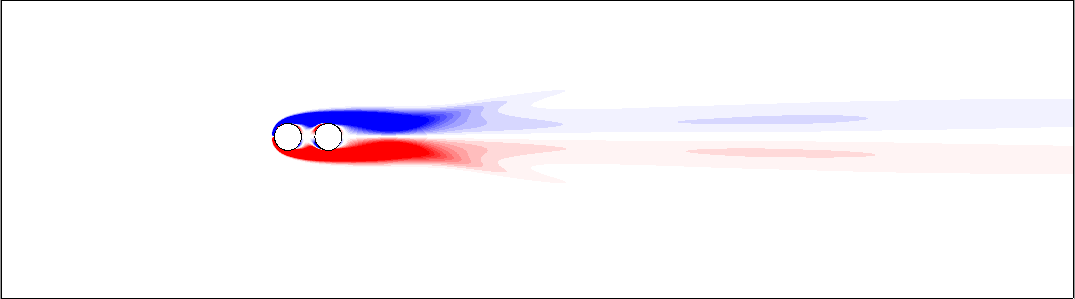}&
\hspace*{-0.02cm}\includegraphics[width=.195\textwidth, trim={7 1 1 1},clip]{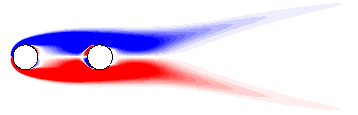}&
\hspace*{-0.055cm}\includegraphics[width=.195\textwidth, trim={10 1 240 1},clip]{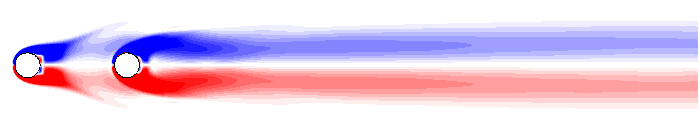}&
\hspace*{-0.34cm}\includegraphics[width=.265\textwidth, trim={70 5 290 7},clip]{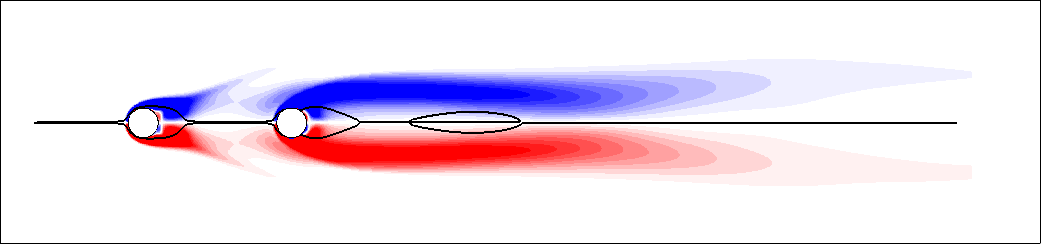}\\

\end{tabular}

  \begin{picture}(0,0)

    \put(123,53){\line(0,1){27}}
    \put(165,53){\line(0,1){27}}
    \put(151,56){\vector(1,0){14}}
    \put(137,56){\vector(-1,0){14}}
    \put(138,54){\text{${L_R}$}}

    \put(123,8){\line(0,1){27}}
    \put(165,8){\line(0,1){27}}
    \put(151,11){\vector(1,0){14}}
    \put(137,11){\vector(-1,0){14}}
    \put(138,9){\text{${L_R}$}}

  \end{picture}

  \caption{\label{fig:Instantaneous}Instantaneous (top) and mean
    (bottom) flow of the two-cylinder system for $p=1.5, 3.0, 4.0$ and
    $5.0$. Red/blue contours mark positive/negative vorticity in the
    range $\Omega D/U = \pm 1$. Solid lines on the mean flow for
    $p = 5$ mark separating streamlines.}
\end{figure}

The base case to be considered is the two-cylinder system. Figure
\ref{fig:Instantaneous} shows examples of the flow generated for this
system as the pitch $p$ is varied. Instantaneous and time-mean images
of the flow are visualised using contours of vorticity.

It is clear that the presence of the second cylinder in the wake of
the first has a strong influence on the overall flow. At short $p$,
the inherent wake dynamics and vortex shedding from the front cylinder
are completely suppressed. At longer $p$, vortex shedding occurs but
the frequency of this shedding is weakly impacted. In both these
situations, there is a feedback mechanism that communicates the
presence of the second cylinder and the associated flow modification
back to the first cylinder. The flow at a given distance downstream is
also a strong function of the position of the second cylinder - for
longer $p$, (i.e., $p \geqslant 4.6$), where vortices shed from the
first cylinder impinge on the second cylinder, a two-row vortex
structure is formed downstream as shown for the example at $p=5.0$ in
figure \ref{fig:Instantaneous} and observed in previous studies
\citep{wang_2010, carmo_2010}. This flow is distinct from the classic
B\'{e}nard-Von \karman\ vortex street in the wake of a single
cylinder.

This two-row structure persists for up to $8D$ downstream of
the second cylinder where secondary instabilities begin to appear,
which coincides with the end of an elongated recirculation region in
the mean flow.

There is a transition state that occurs over a small range of $p$ that
precludes the appearance of this two-row structure. From the onset of
vortex shedding in the gap between the cylinders at $p\simeq 3.8$,
until the onset of the two-row structure at $p=4.6$, a single row of
vortex shedding with higher frequency can be observed in the wake of
the second cylinder as shown for $p=4.0$ in figure
\ref{fig:Instantaneous}.

Also marked on figure \ref{fig:Instantaneous} for the longer-pitch
cases is the mean recirculation length \convectivelength, defined as
the end of the mean flow recirculation region, indicated in the figure
by the separating streamlines. Comparing the mean and instantaneous
images, it can be seen that this length coincides with the point of
the breakdown of the two-row structure, as evidenced by the beginning
of some waviness of the zero-vorticity (white) layer separating the
positive and negative vortices on either side of the wake.

\subsection{Local stability analysis of the two-cylinder wake}
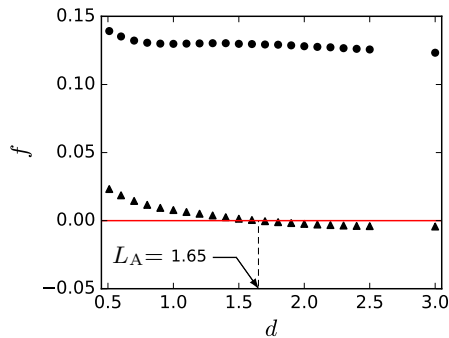
\begin{figure}
  \centerline{
  \includegraphics[width=.45\textwidth, trim={12 0 0 10},clip]{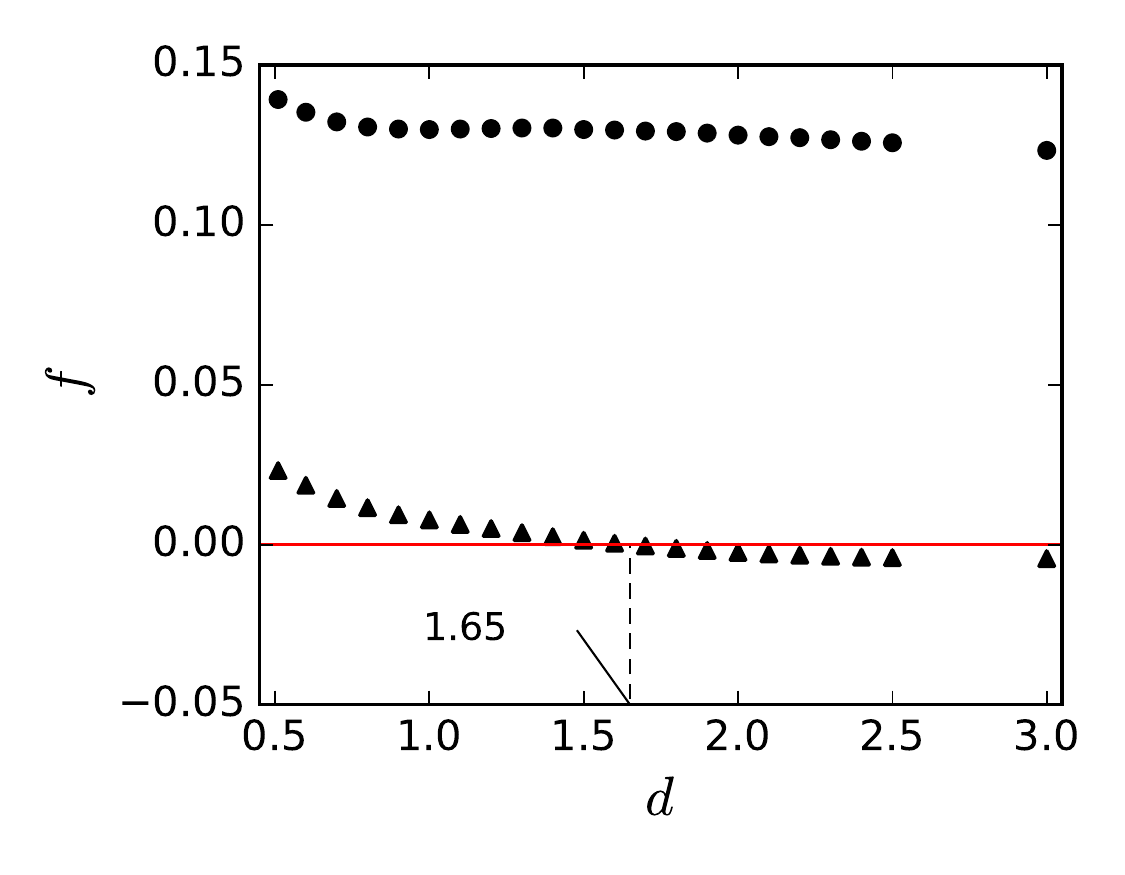}}

  \caption{\label{fig:frequencies}Calculated real ($\bullet$) and
    imaginary ($\blacktriangle$) frequencies from local instability
    analysis for the two-cylinder system with $p=5.0$ as a function of
    the distance from the centre of the second cylinder ($d$). The
    value \abslength\ marks the distance below which the flow is
    absolutely unstable.}

\begin{picture}(0,0)
\put(193.8,83.9){\drawline(-1,0)(-10,0)}
\put(192.8,83.75){\vector(1,-1.4){8.5}}
\put(146,81.9){\abslength = }
\end{picture}
\end{figure}

To further investigate this convective structure, we have performed a
local stability analysis of the mean flow. Figure
\ref{fig:frequencies} shows the result of the local stability analysis
for the case of the two-cylinder system with $p = 5.0$. The imaginary
component is always close to zero, and is only slightly positive in a
small region within $1.65D$ of the body. For distances further
downstream, the imaginary component is always slightly negative. This
indicates a small region that is locally absolutely unstable - here
dubbed \abslength, for the length of the absolutely unstable region -
followed by a large region that is convectively unstable extending at
least to \convectivelength. Note the presence of this small region
that is absolutely unstable does not guarantee any influence on the
global dynamics - it is a necessary but not sufficient condition for
the emergence of a global instability \citep{chomaz2005global}. We
propose that the nonlinear saturation process in the wake of the
second body can be understood in, or decomposed into, two stages; the
presence of the second body in the \karman\ vortex street of the first
body causes a mean flow that is generally convectively unstable, and
this convectively unstable mean flow then responds at the frequency of
the forcing provided by the vortex shedding from the upstream
cylinder, resulting in the two-row structure. The mean flow structure
and therefore the characteristic frequency \freqcc\ is almost
unchanged in the wake up to the breakdown of the two-row structure, or
\convectivelength.

Absolute instabilities have the potential to act as a feedback
mechanism. Disturbances introduced by placing a body in regions that
are absolutely unstable can be amplified and fed upstream and
downstream. Convective instabilities can only carry disturbances
downstream, therefore bodies introduced into a convectively unstable
region should not be felt upstream. Therefore, the broadcasting region
should correspond to the region which is absolutely unstable, and the
cloaking region at least to a region which is convectively unstable.

\begin{figure}
  \centering
\begin{tabular}{>{\centering\arraybackslash}m{1.7cm}m{6cm}m{5cm}}

    &
  \hspace*{0cm}Third body: cylinder&
  \hspace*{0cm}Third body: plate\\

$\pthree\rightarrow\infty$  &
\includegraphics[width=.35\textwidth, trim={350 75 180 20},clip]{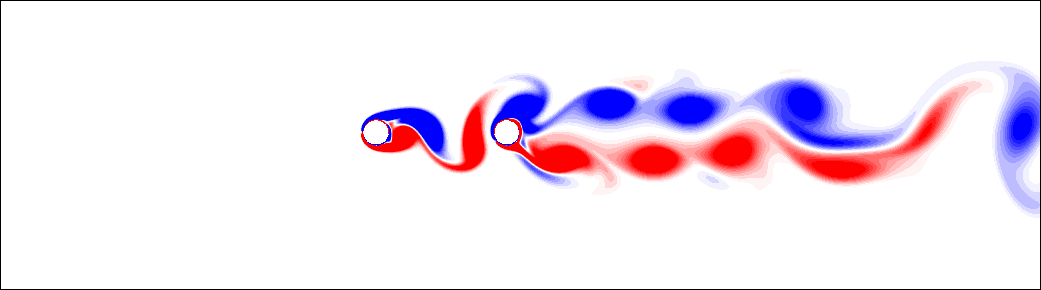}&
\includegraphics[width=.35\textwidth, trim={350 75 180 20},clip]{images/inst_p50.png}\\

$\pthree=7.0$  &
\includegraphics[width=.35\textwidth, trim={150 85 300 80},clip]{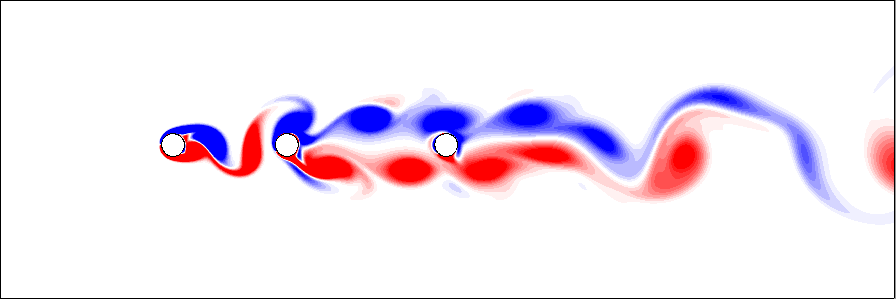}&
\includegraphics[width=.35\textwidth, trim={150 85 300 80},clip]{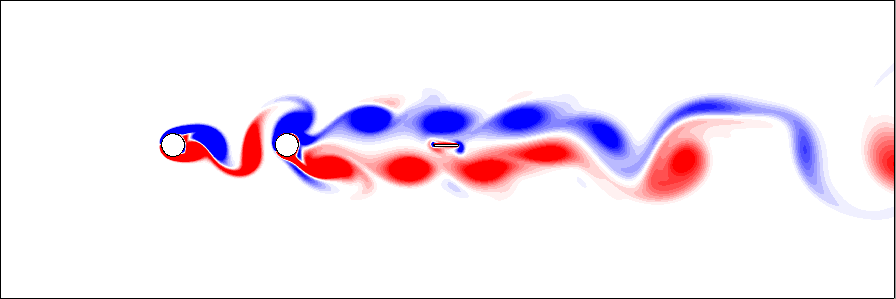}\\

$\pthree=6.0$  &
\includegraphics[width=.35\textwidth, trim={150 85 300 80},clip]{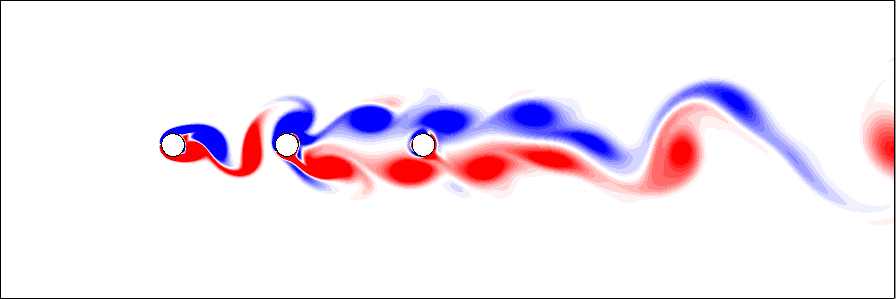}&
\includegraphics[width=.35\textwidth, trim={150 85 300 80},clip]{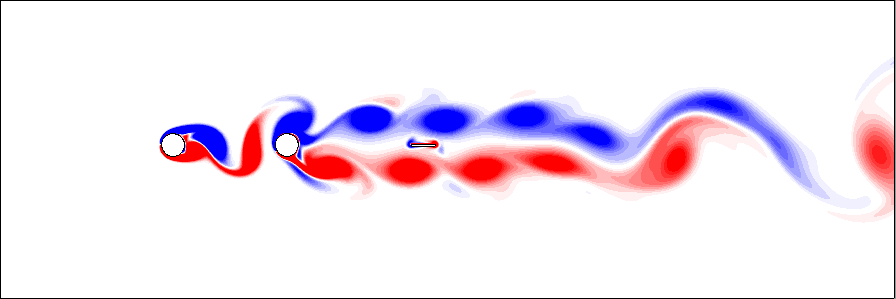}\\

$\pthree=5.0$  &
\includegraphics[width=.358\textwidth, trim={150 85 285 80},clip]{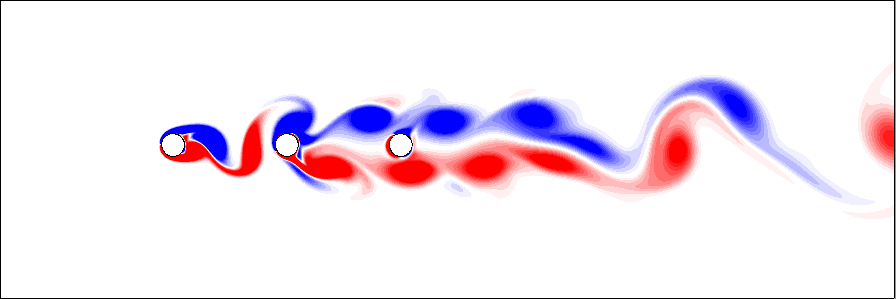}&
\includegraphics[width=.358\textwidth, trim={150 85 285 80},clip]{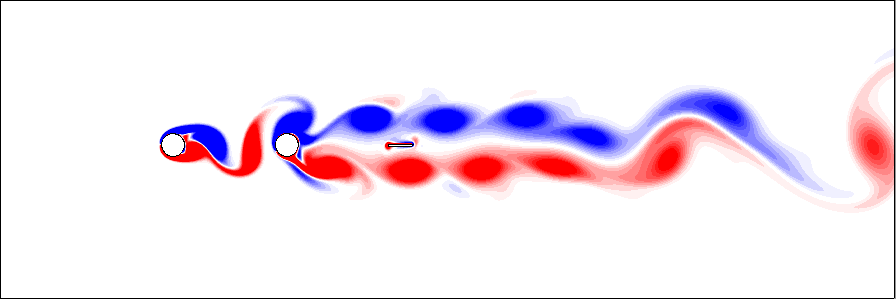}\\

$\pthree=4.0$  &
\includegraphics[width=.35\textwidth, trim={150 85 300 80},clip]{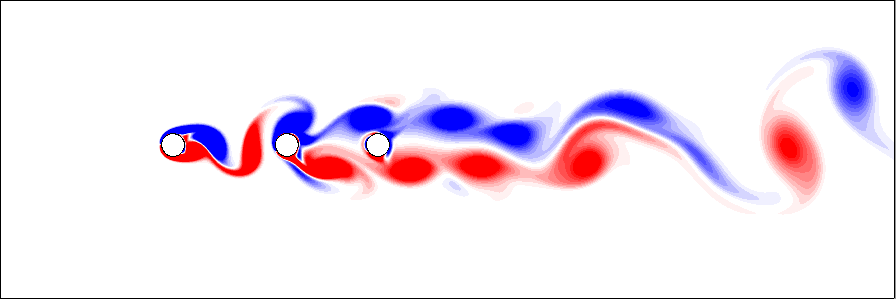}&
\includegraphics[width=.35\textwidth, trim={150 85 300 80},clip]{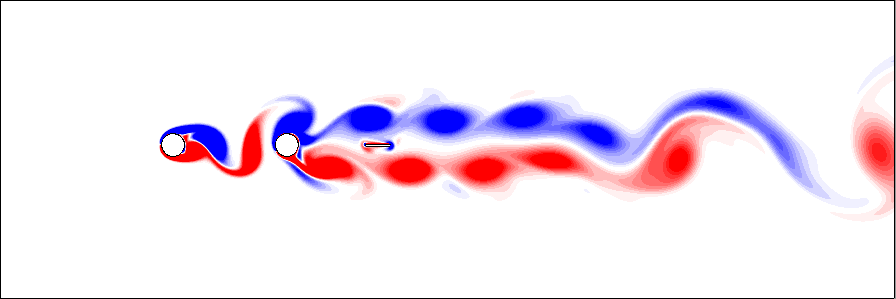}\\

$\pthree=3.0$  &
\includegraphics[width=.35\textwidth, trim={150 85 300 80},clip]{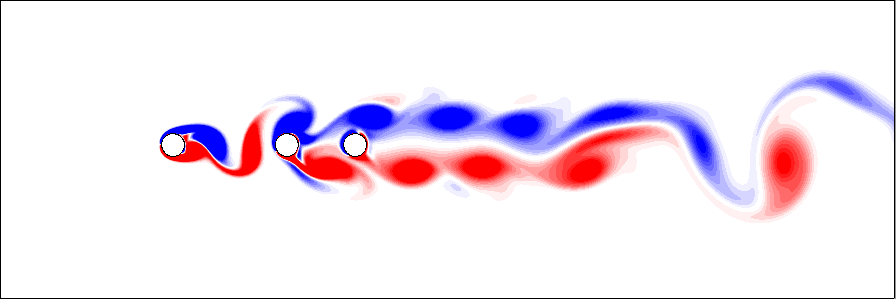}&
\includegraphics[width=.35\textwidth, trim={150 85 300 80},clip]{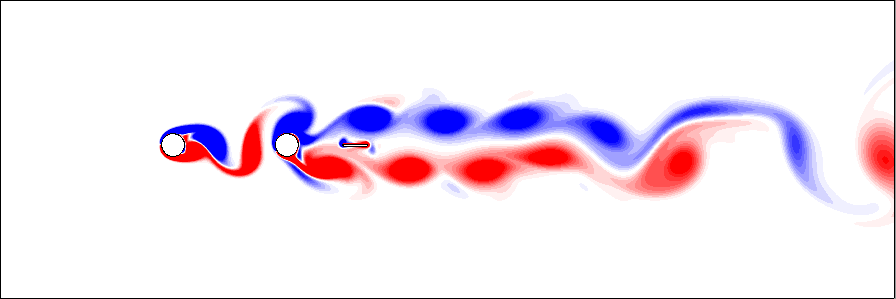}\\

$\pthree=2.0$  &
\includegraphics[width=.35\textwidth, trim={150 85 300 80},clip]{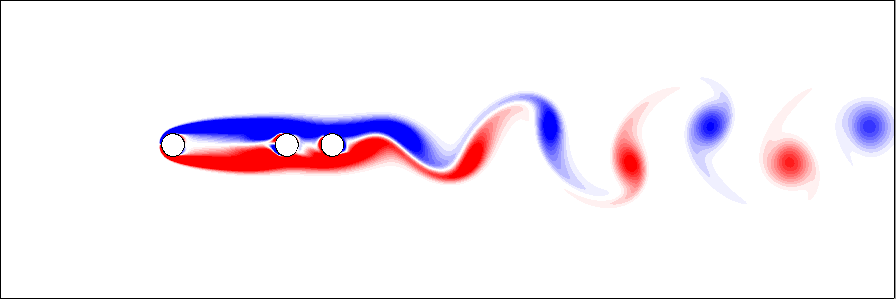}&
\includegraphics[width=.35\textwidth, trim={150 85 300 80},clip]{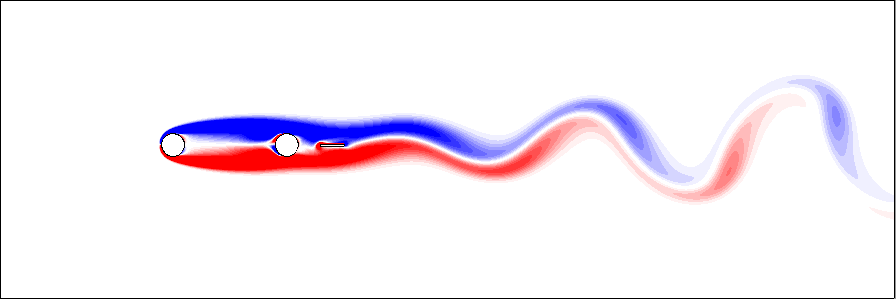}\\

\end{tabular}

\begin{picture}(0,0)
  \multiput(97,33)(0,7){4}{\line(0,1){4}}
  \multiput(-79.9,33)(0,7){4}{\line(0,1){4}}
\end{picture}
 
\caption{\label{fig:diff_con_len}Instantaneous flow visualizations for
  three-body systems. The first two bodies are cylinders separated by
  $p=5.0$ while a third body (cylinder or plate) is placed different
  distances \pthree\ from the second cylinder. All images are at
  maximum lift on the second cylinder. Dashed lines mark the end of
  the absolutely unstable region \abslength\ in the two-cylinder setup.}
\end{figure}

\subsection{Introduction of a third body to the row}
Figure \ref{fig:diff_con_len} clearly demonstrates these broadcasting
and cloaking phenomena and their correspondence to the absolutely and
convectively unstable regions, showing results from a series of
simulations with a third body placed in the two-row structure
wake. The distance between the first two cylinders is kept constant at
$p = 5.0$, however the distance between the second and third body
\pthree\ is varied independently. The first column shows flows where
the third body is a circular cylinder identical to the others, in the
second column the third body is a small flat plate of length $D$ and
thickness $0.022D$, with rounded semi-circular ends. The first image
at the top of both columns shows the two cylinder case and the
subsequent images show \pthree\ becoming gradually shorter. All the
images are captured at an instant where the lift force on the second
cylinder is maximum, and hence at the same phase of the vortex
shedding cycle on this cylinder.

The results are striking - for values of $\pthree - D/2 > \abslength$
(we consider the most upstream point of the body as the first point to
cause perturbation in the flow), the presence of the third body in
either geometry has essentially no impact on the flow structure. The
convectively unstable shear layers simply pass over the third body. No
disturbance is felt upstream as expected for a convectively unstable
flow. Further, as the flow on the wake centreline is close to
stationary, the presence of the third body that forces the flow
velocity to exactly zero at its surface does not introduce any
significant disturbance downstream. Comparing images down each column
as \pthree\ is reduced shows that the position of the first three
vortex pairs in the wake is unaffected by the introduction of the
third body, and comparing across rows shows the structure is also
insensitive to the shape of the third body.

Moving the third body to a distance $\pthree - D/2< \abslength$ from the
second body has a profound impact on the flow as shown in the very
bottom images of figure \ref{fig:diff_con_len} where $\pthree =
2.0$. This places the most upstream point of the body $1.5D$ from the
centre of the second body, or inside the absolutely unstable
region. The vortex shedding from the first cylinder is completely
suppressed, and with the loss of the impinging vortices on the second
cylinder the two-row structure is also destroyed. The flow transits to
a new global state with a lower frequency vortex shedding in the wake
of the array as evidenced by the longer wavelength vortex shedding
apparent in the bottom images of figure \ref{fig:diff_con_len}.

The results shown in figure \ref{fig:diff_con_len} demonstrate the
cloaking and broadcasting phenomena. They show that these phenomena
occur in a way that is independent of the details of the third body
geometry, but is strongly dependent on the third body position. This
strong dependence on position appears to correlate with the existence of
absolutely and convectively unstable regions of the mean flow, and the
boundary between these regions provides a threshold for the position
of the third body to be broadcast (trigger a global flow change) or to
be cloaked (have almost zero global impact).



\begin{figure}
  \centering
\begin{tabular}{>{\centering\arraybackslash}m{5cm}m{5cm}m{5cm}}

\hspace*{-1cm}$\pthree=2.0$  &
 \hspace*{0.89cm}$\pthree=2.1$  &
$\pthree=2.2$ \\

\hspace*{-1cm}\includegraphics[width=.26\textwidth, trim={150 70 240 55},clip]{images/flow_vis_c_1.png}&
\hspace*{-0.3cm}\includegraphics[width=.28\textwidth, trim={20 70 300 65},clip]{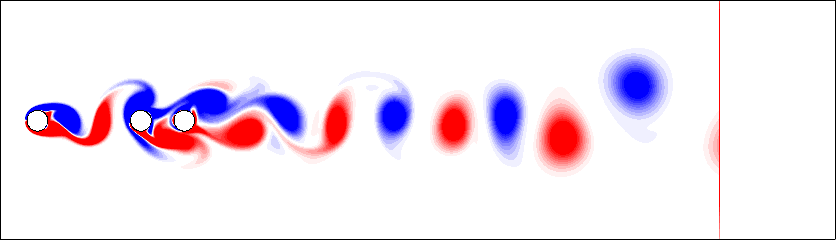} &
\hspace*{-1cm}\includegraphics[width=.28\textwidth, trim={20 50 300 45},clip]{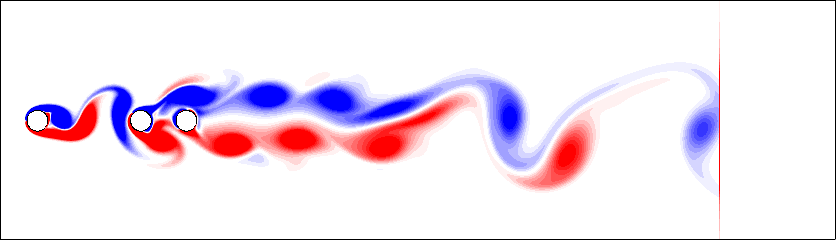} \\

\end{tabular}
 
\caption{\label{fig:transition}Instantaneous flow visualizations for
  three-body systems in the broadcasting, transition and cloaking
  states for $p=5.0$ and $\pthree =2.0,2.1$ and $2.2$, respectively. All
  images are shown at an instant of maximum lift on the second
  cylinder.}
\end{figure}

To check the accuracy of the presented data and sensitivity to the
location of the third cylinder, we performed simulations at smaller
increments of $\pthree$ in the range of
$\pthree - D/2 \approx \abslength$. Flow visualizations for these
cases are illustrated in figure~\ref{fig:transition}.




These images show that in a very small range of positions, where the
most upstream point of the third body is around the $L_A=1.65$ found
from local stability analysis (see figure~\ref{fig:frequencies}) the
broadcasting phenomenon shifts to a cloaking phenomenon through a
small transition region. We note the appearance of this transition
flow, which consists of a single row of vortices shed into the
eventual wake (figure~\ref{fig:transition}, $\pthree =2.1$) is similar
to the transition state of two-cylinder system
(figure~\ref{fig:Instantaneous}, $p =4.0$) before the appearance of
the two-row structure. In the three-cylinder case, there is a vortex
formation behind the most upstream cylinder, a short region of the
two-row structure behind the second body, followed by a single-row
vortex formation with a lower frequency.

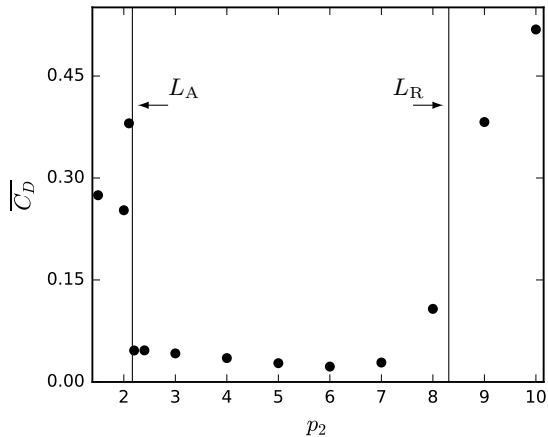
\begin{figure}\centering
  \setlength{\unitlength}{\textwidth}
  \begin{picture}(1,0.42)\centering
    \put(0.225,0){\includegraphics[width=0.55\textwidth, trim={13.5 10 0 10},clip]{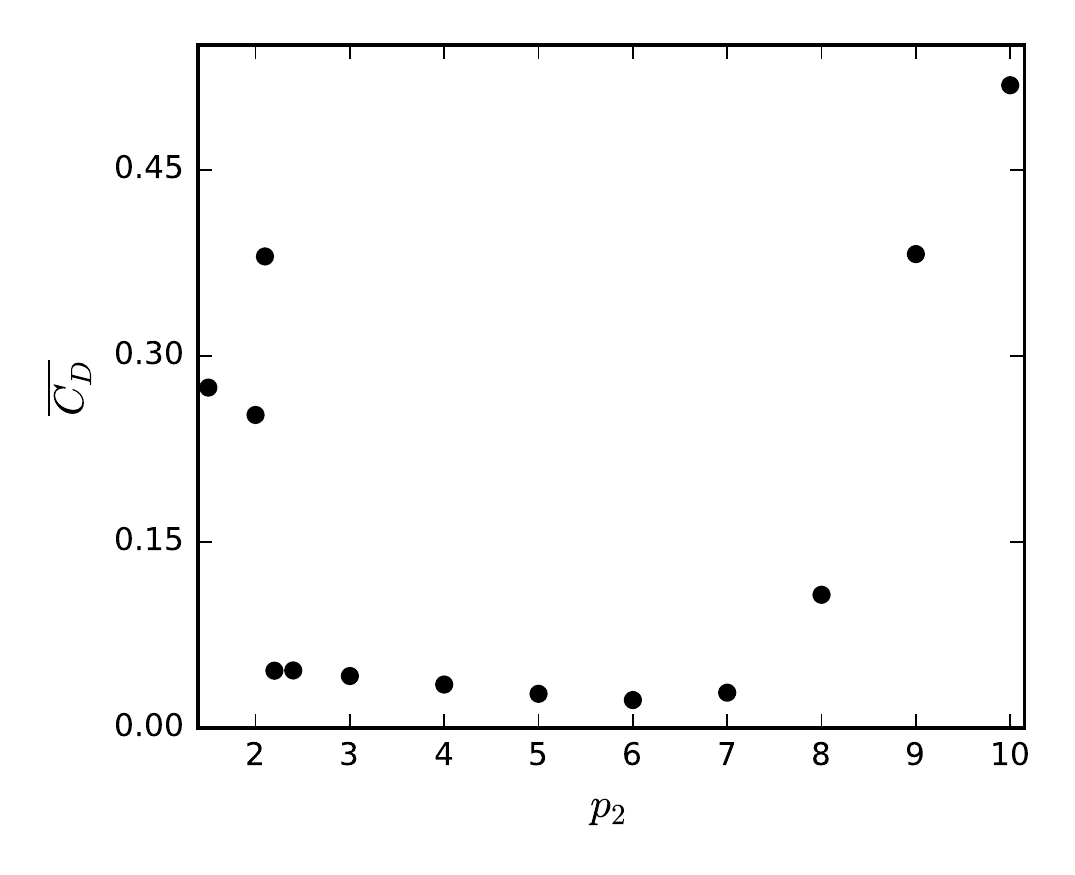}}
    \put(0.345,0.06){\line(0,1){0.365}}
    \put(0.655,0.06){\line(0,1){0.365}}

    \put(0.6,0.34){\convectivelength}
    \put(0.38,0.34){\abslength}

    \put(0.38,0.33){\vector(-1,0){0.03}}
    \put(0.62,0.33){\vector(1,0){0.03}}
  \end{picture}
  \caption{\label{fig:drag_plot}The plot of $\overline{C_D}$ as a
    function of $\pthree$ for a three-cylinder system with consistent
    $p=5.0$ and different $\pthree$. Vertical solid lines mark
    locations of \abslength\ calculated from the local stability
    analysis, and \convectivelength\ calculated from the mean flow
    streamlines.}
\end{figure}

Figure~\ref{fig:drag_plot} shows the mean drag coefficient
$\overline{C_D}$ on the third cylinder in different positions behind
the second cylinder $\pthree$ with consistent $p=5.0$.  When
$\convectivelength > \pthree -D/2> \abslength$ the mean drag decreases
to values very close to zero. The transition from a broadcasting to
cloaking state is concurrent with a sudden drop in the drag. Beyond
the \convectivelength\ region and the end of the two-row structure,
$C_D$ increases again.

\begin{figure}
  \centering
  \includegraphics[width=.5\textwidth, trim={12 10 10 10},clip]{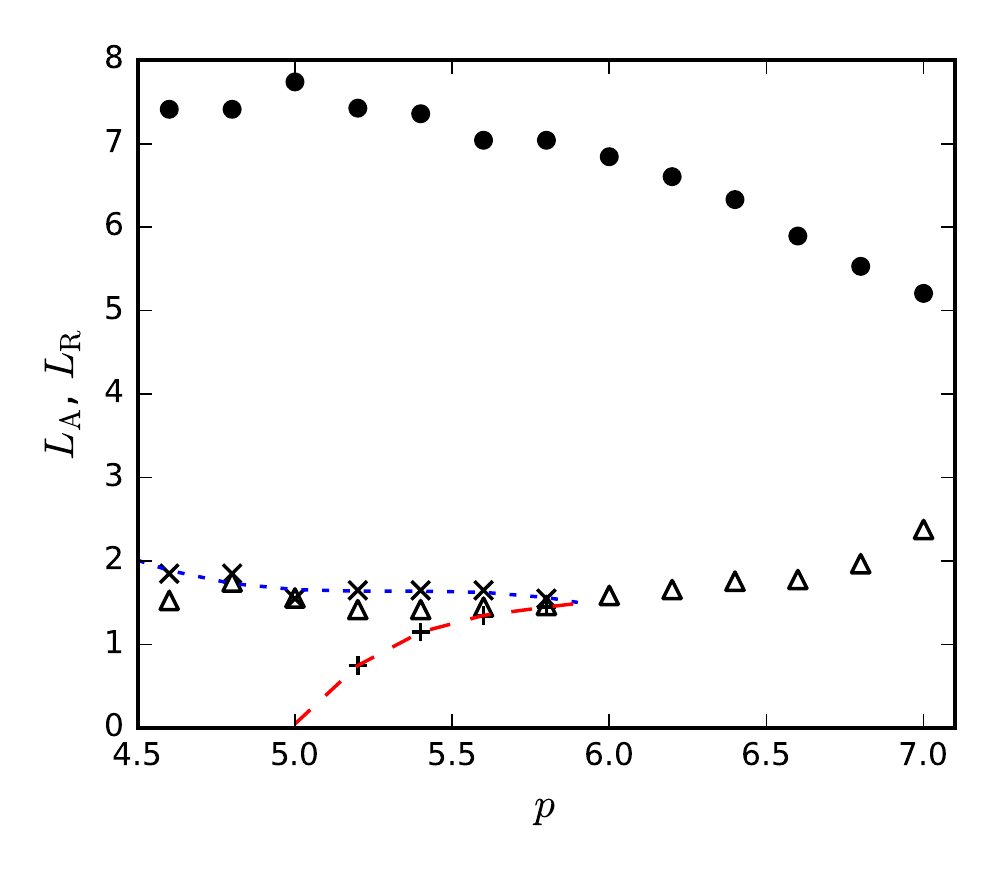}
  \caption{\label{fig:L_cPlot}Mean recirculation length
    \convectivelength\ ($\bullet$) and the length of the locally
    absolutely unstable region \abslength\ (${\tiny\triangle}$) for
    two-cylinder system as a function of $p$ using the local stability
    analysis. The \abslength\ data are evaluated by simulation are
    presented with $\times$ and $+$ for upper and lower boundaries,
    respectively. The dashed lines connecting these data are cubics of
    best fit in the form of
    $L_{A_{max}} = -0.694p^3 + 11.071p^2 - 58.829p+ 105.826$ and
    $L_{A_{min}} = 2.083p^3 - 36.25p^2 + 210.666p - 407.45$.}
\end{figure}

The results presented for $p=5.0$ can be generalised to a range of
$p$. Figure \ref{fig:L_cPlot} plots both \abslength\ and
\convectivelength\ as a function of the pitch $p$, where both lengths
are relative to the centre of the second cylinder, for the range where
the two-row structure is observed ($p \geqslant 4.6$) from the
two-cylinder system. This plot therefore provides a prediction of the
range of the broadcasting and cloaking regions from the local
stability analysis - the broadcasting region can occupy the wake of
the second cylinder up to \abslength, and the cloaking region occupies
the wake for the second cylinder from \abslength\ up to at least
\convectivelength.

Also shown on figure \ref{fig:L_cPlot} are the values of the threshold
between the broadcasting and cloaking regions found simply by running
simulations with a third body placed at a varying distance
downstream. The body position plotted is the distance from the centre
of the second cylinder to the most upstream point of the third
cylinder as it is theorised that this is the first point to perturb
the absolutely unstable region. The method for finding this threshold
is illustrated for the example case at $p=5.0$ above in figures
\ref{fig:diff_con_len} and \ref{fig:transition}. These direct
measurements using three bodies 
follow the predicted values using two bodies very closely for
$p \leq 5.8$.  For $p > 5.8$ the distance between the first two
cylinders is so large that any feedback mechanism between the
absolutely unstable region and the upstream cylinder is broken, and
the presence of the third body in \abslength\ ceases to cause a global
change of the flow. The two-row structure persists, with the
closely-spaced second and third cylinders essentially behaving as a
single elongated body.

\subsection{Establishment of a lower bound for the body position that
  can trigger the global change}
\begin{figure}
  \centering
  \begin{tabular}{>{\centering\arraybackslash}m{2.0cm}m{2.0cm}m{2.0cm}m{2.0cm}m{2.0cm}m{2.0cm} }

\hspace*{0cm}$\pthree=1.2$  &
\hspace*{0.4cm}$\pthree=1.3$ &
\hspace*{0.4cm}$\pthree=1.7$ &
\hspace*{0.4cm}$\pthree=2.1$ &
\hspace*{0.4cm}$\pthree=2.2$ &
\hspace*{0.4cm}$\pthree=2.3$ \\

\hspace*{-0.2cm}\includegraphics[width=0.17\textwidth, trim={20 50 260 50},clip]{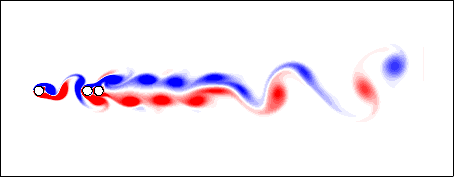}&
\hspace*{-0.2cm}\includegraphics[width=0.17\textwidth, trim={20 50 260 50},clip]{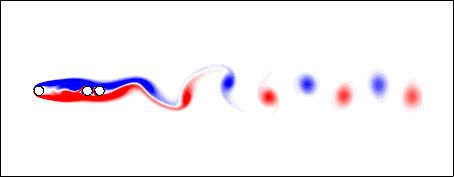}&
\hspace*{-0.2cm}\includegraphics[width=0.17\textwidth, trim={20 50 260 50},clip]{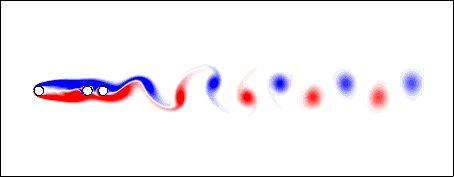} &
\hspace*{-0.2cm}\includegraphics[width=0.17\textwidth, trim={20 50 260 50},clip]{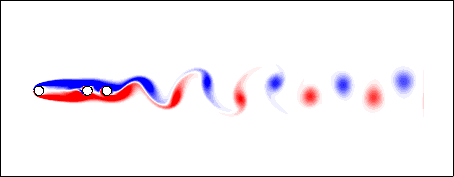}&
\hspace*{-0.2cm}\includegraphics[width=0.17\textwidth, trim={20 50 260 50},clip]{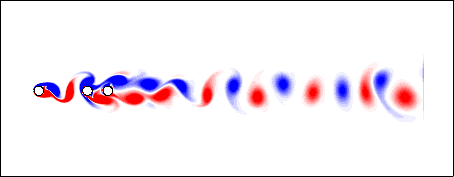} &
\hspace*{-0.2cm}\includegraphics[width=0.17\textwidth, trim={20 50 260 50},clip]{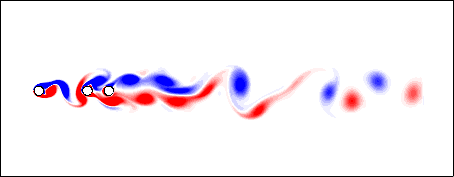} \\

\end{tabular}
 
\caption{\label{fig:FLow_vis_different_states52}Instantaneous flow
  visualization of the flow before, across and after the broadcasting
  region for $p=5.2$ and varying $\pthree=1.2, 1.3, 1.7, 2.1, 2.2$ and
  $2.3$. All the visualizations are presented at the instant of the
  maximum lift on the second body.}
\end{figure}

A phenomenon that is not predicted by the local stability analysis is
the appearance of a lower boundary on the downstream distance of the
third body to trigger the global change in state and
broadcasting. However, the direct measurements show this very
clearly. Figure \ref{fig:FLow_vis_different_states52} shows a series
of instantaneous flow visualisations for increasing $\pthree$ for a
constant $p=5.2$. For very short $\pthree$ (the first value shown is
$\pthree = 1.2$, which gives a gap between the second and third
cylinders of $0.2D$) there is clear vortex shedding in the first gap,
and the wake of the array is clearly in the two-row convective
structure. However, slightly increasing to $\pthree = 1.3$ sees the
vortex shedding in the first gap cease as the global change in state
is triggered and the wake transits to a modified \karman\ wake with
reduced frequency. There is therefore a lower boundary for the
appearance of the global instability around $\pthree = 1.3$ for
$p = 5.2$. This global state remains until $\pthree = 2.2$, when the
most upstream point of the third cylinder leaves the predicted
absolutely unstable region and the flow moves back to the convectively
unstable wake structure, in line with the prediction from the two
cylinder system.

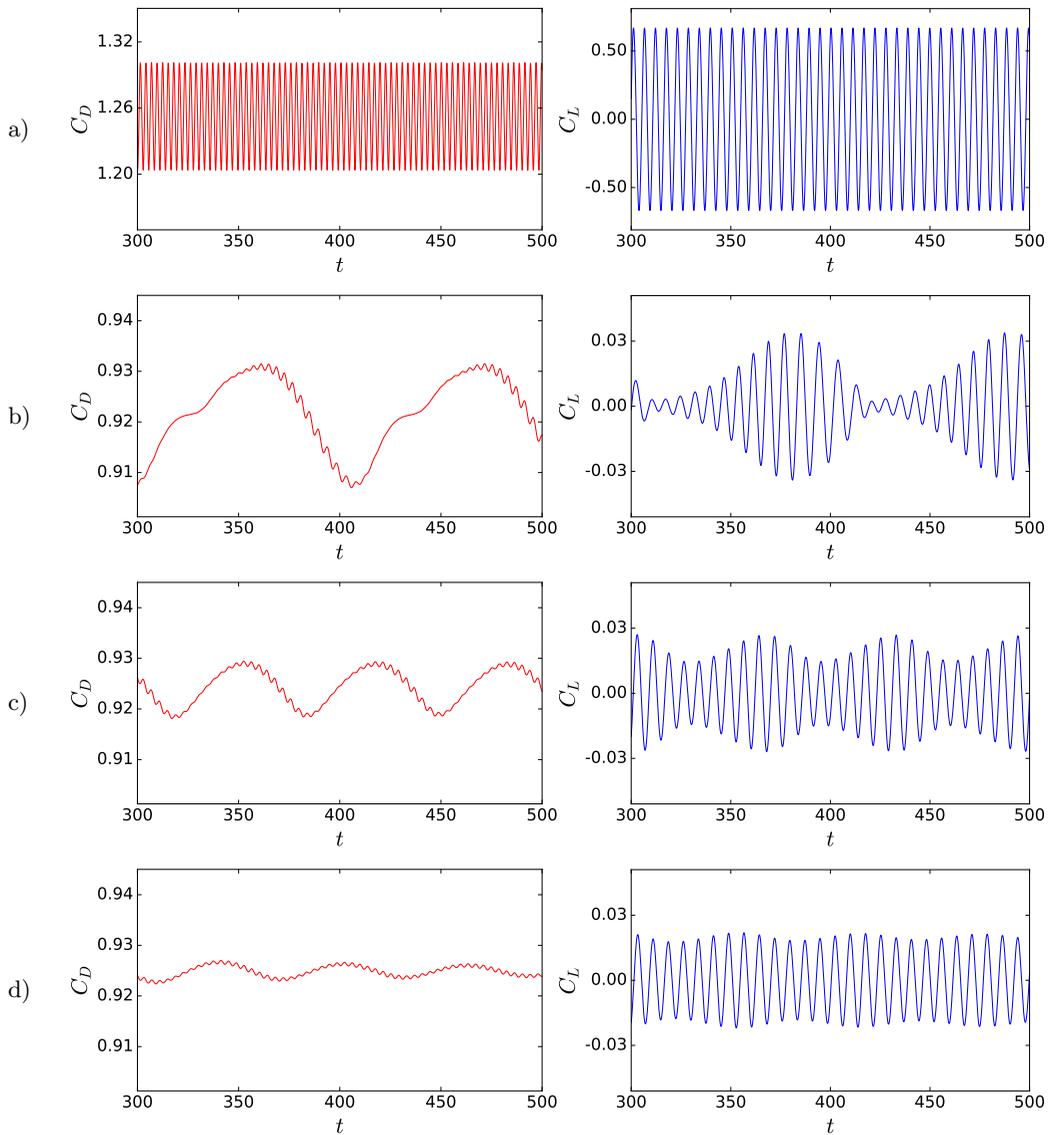
\begin{figure}
 \setlength{\unitlength}{\textwidth}
  \begin{picture}(1,1.2)
\put(-0.01,1.01){a)}
\put(-0.01,0.73){b)}
\put(-0.01,0.45){c)}
\put(-0.01,0.17){d)}
 \put(0.038,0.86){\includegraphics[width=.489\textwidth, trim={0 0 14.485 0},clip]{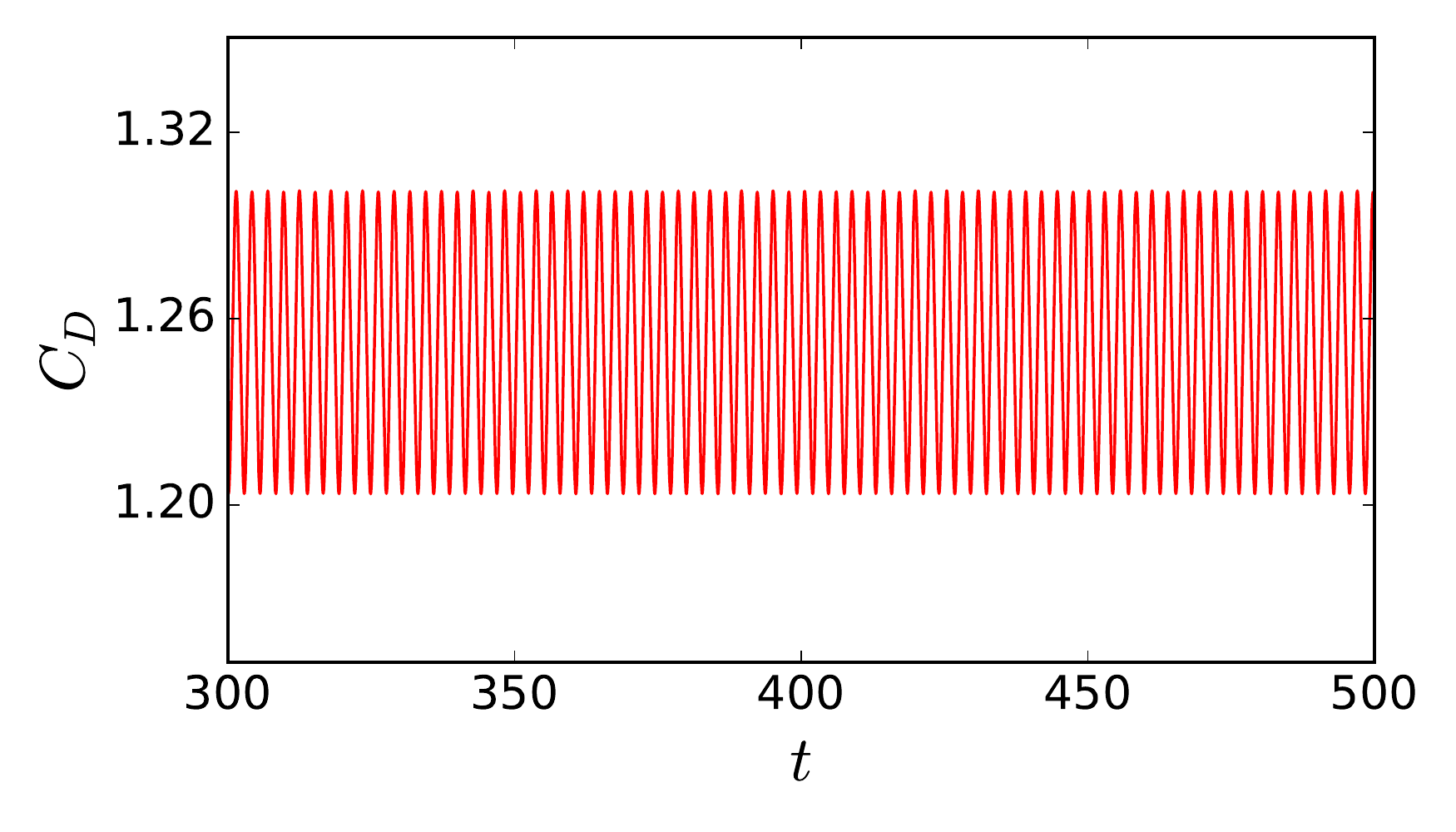}}
 \put(0.53,0.86){\includegraphics[width=.489\textwidth, trim={14 0 0 0},clip]{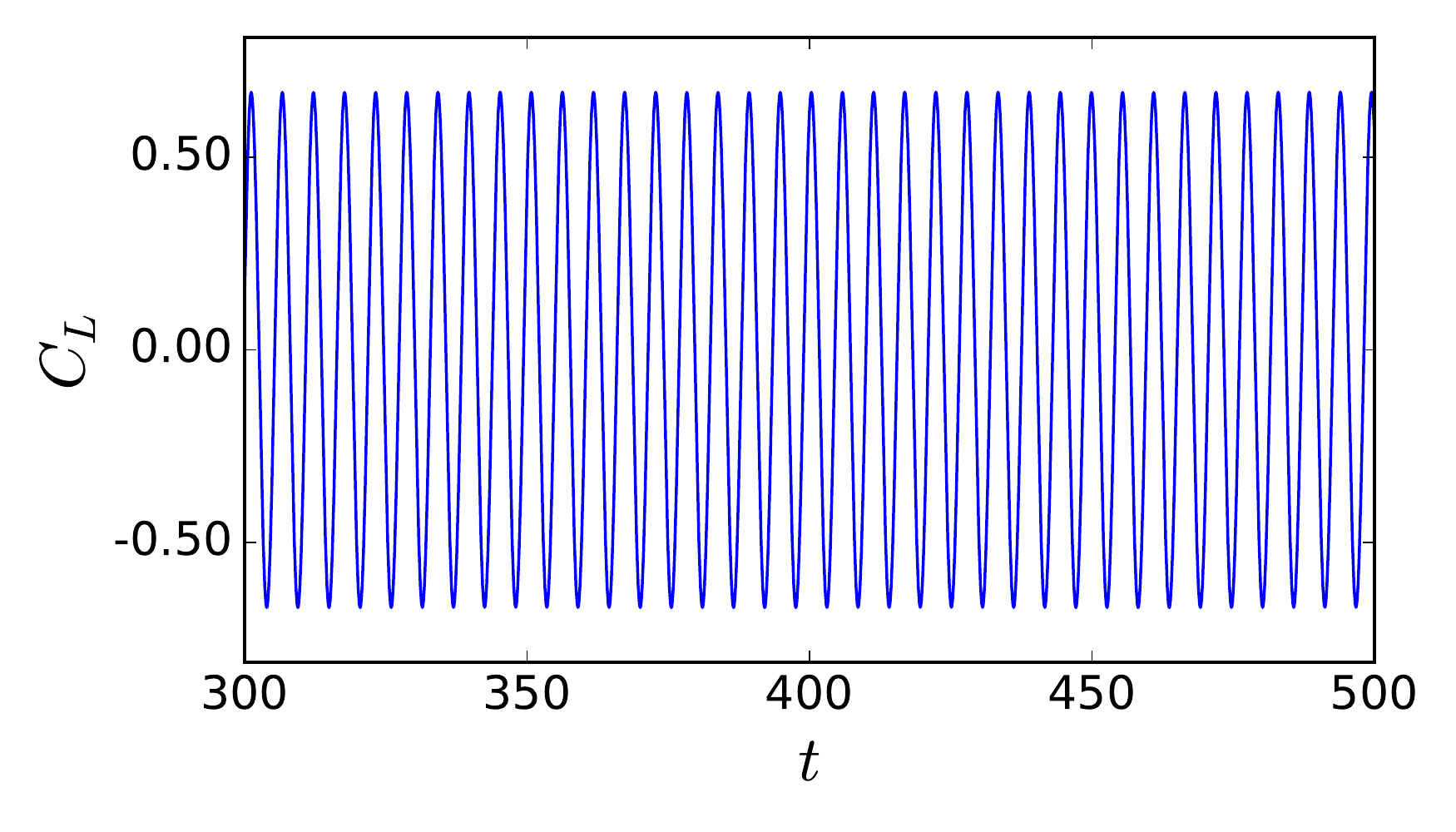}}
 \put(0.038,0.58){\includegraphics[width=.489\textwidth, trim={0 0 14.485 0},clip]{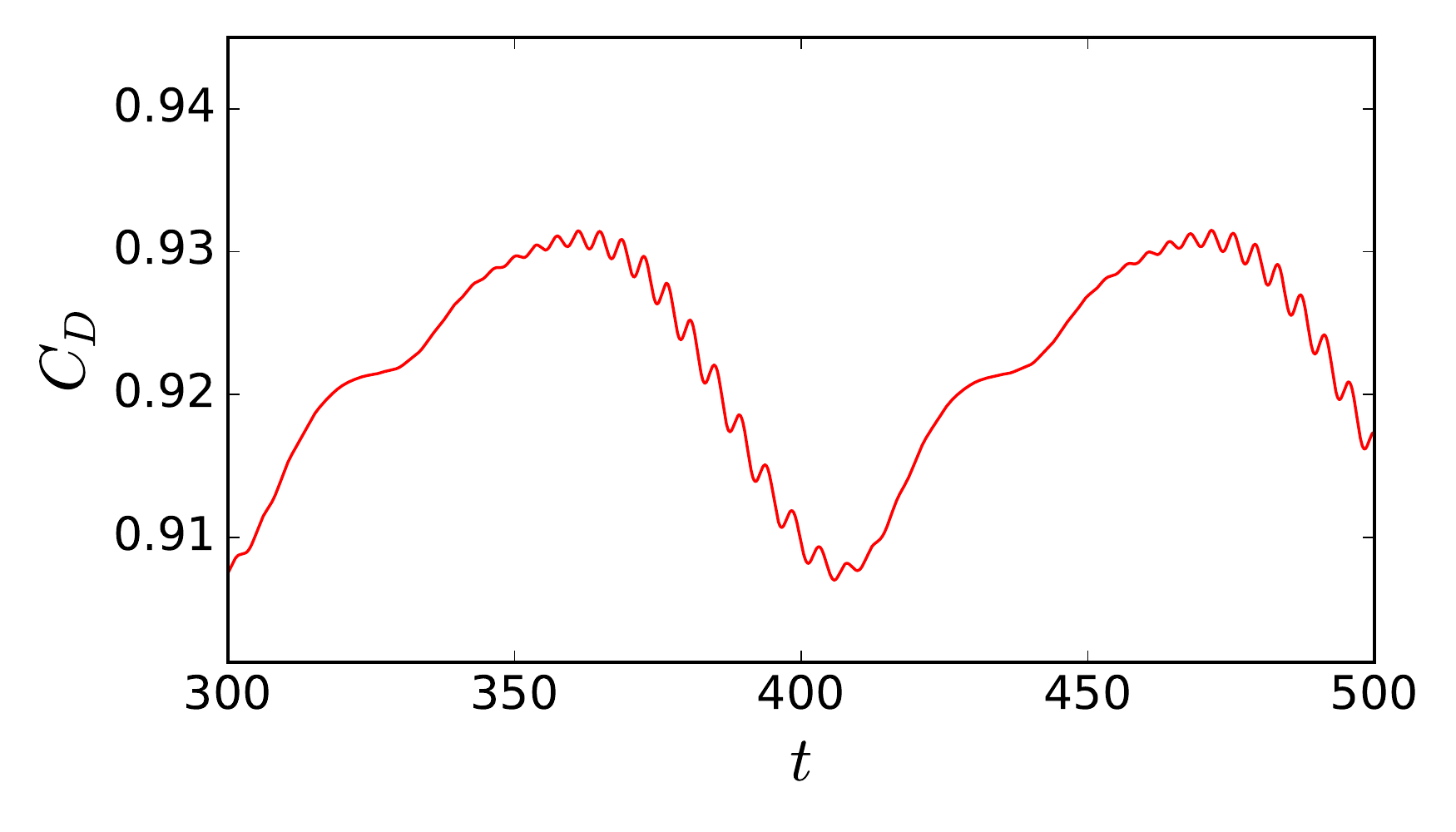}}
 \put(0.53,0.58){\includegraphics[width=.489\textwidth, trim={14 0 0 0},clip]{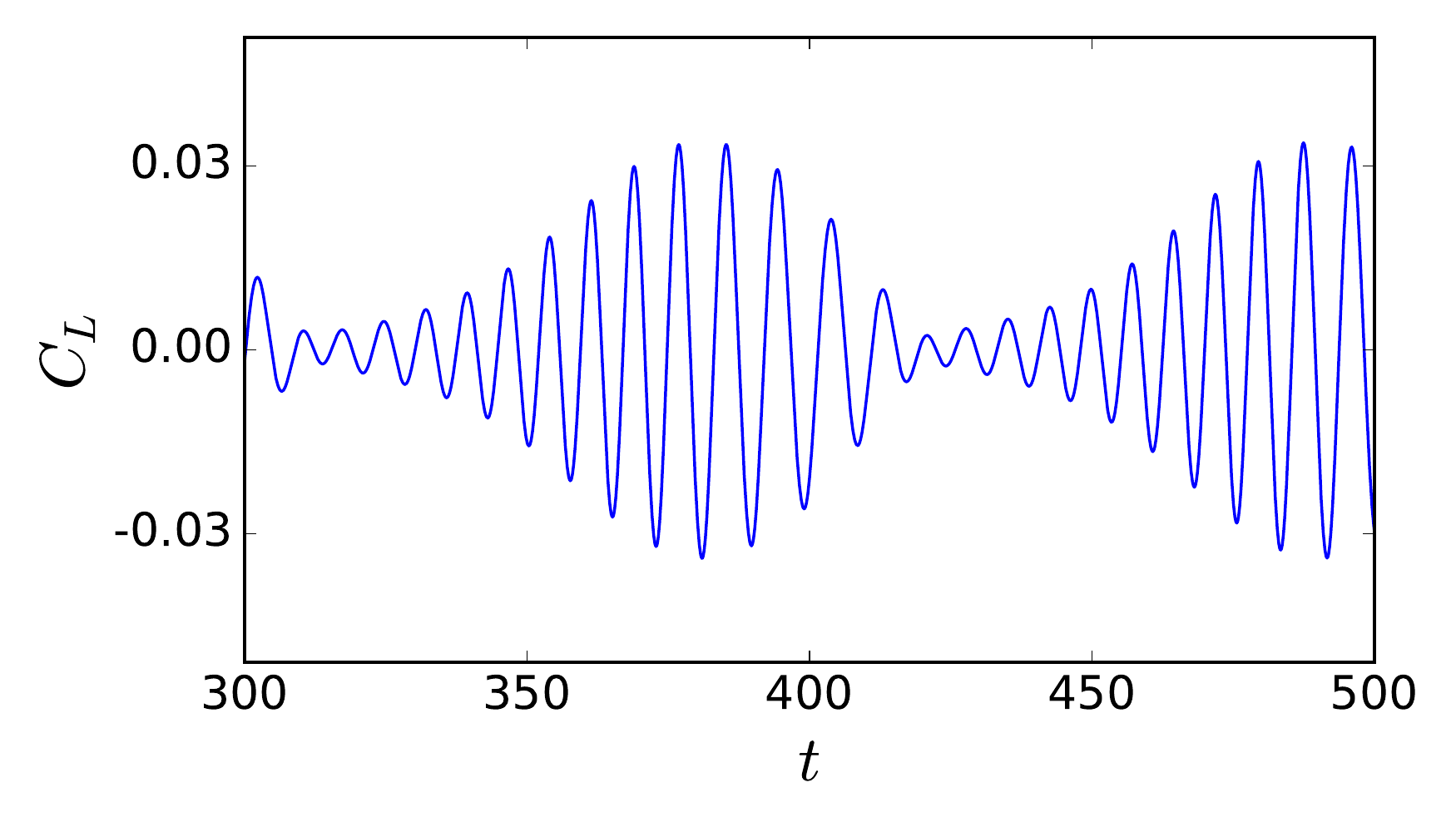}}
 \put(0.038,0.3){\includegraphics[width=.489\textwidth, trim={0 0 14.485 0},clip]{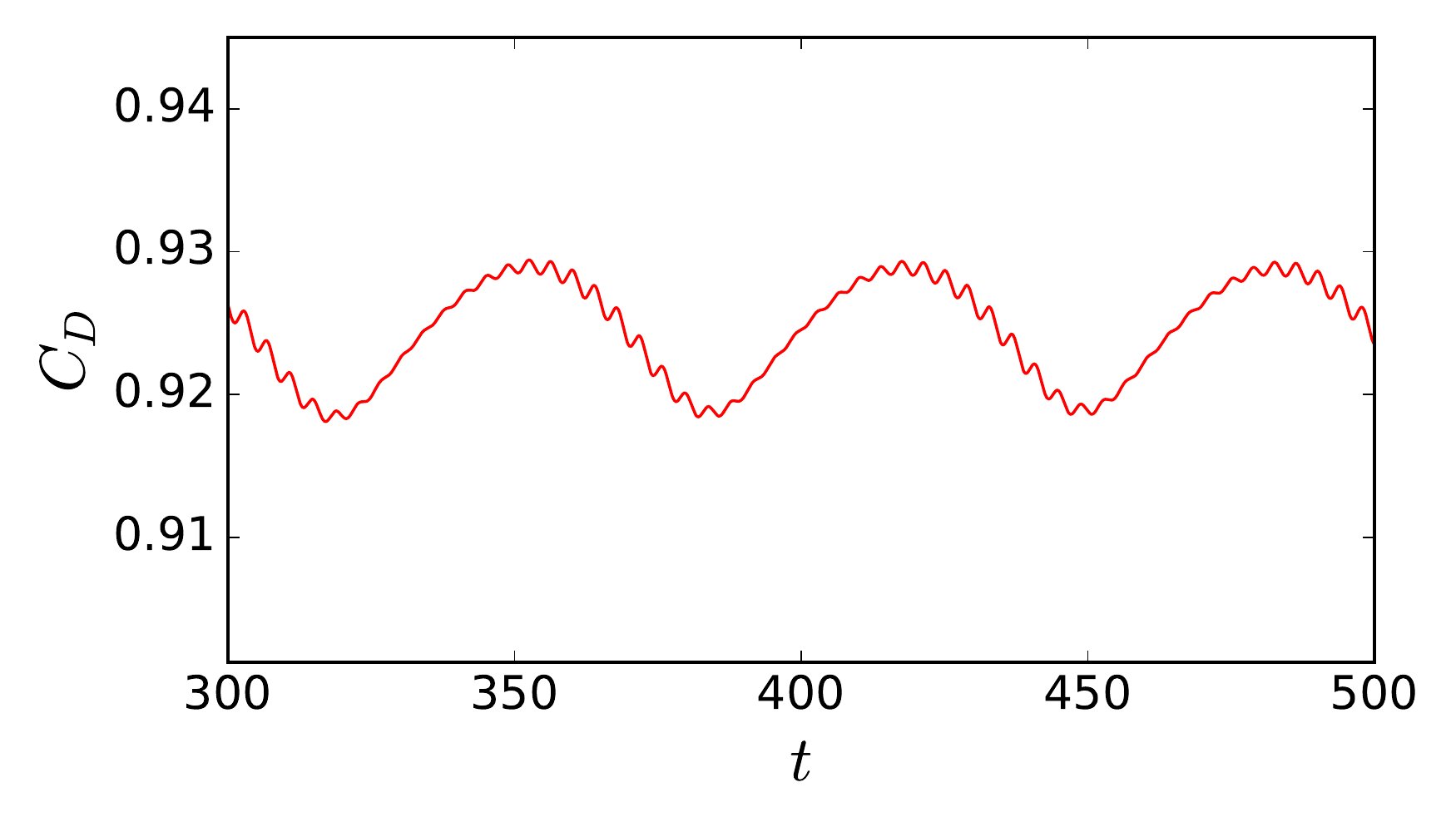}}
 \put(0.53,0.3){\includegraphics[width=.489\textwidth, trim={14 0 0 0},clip]{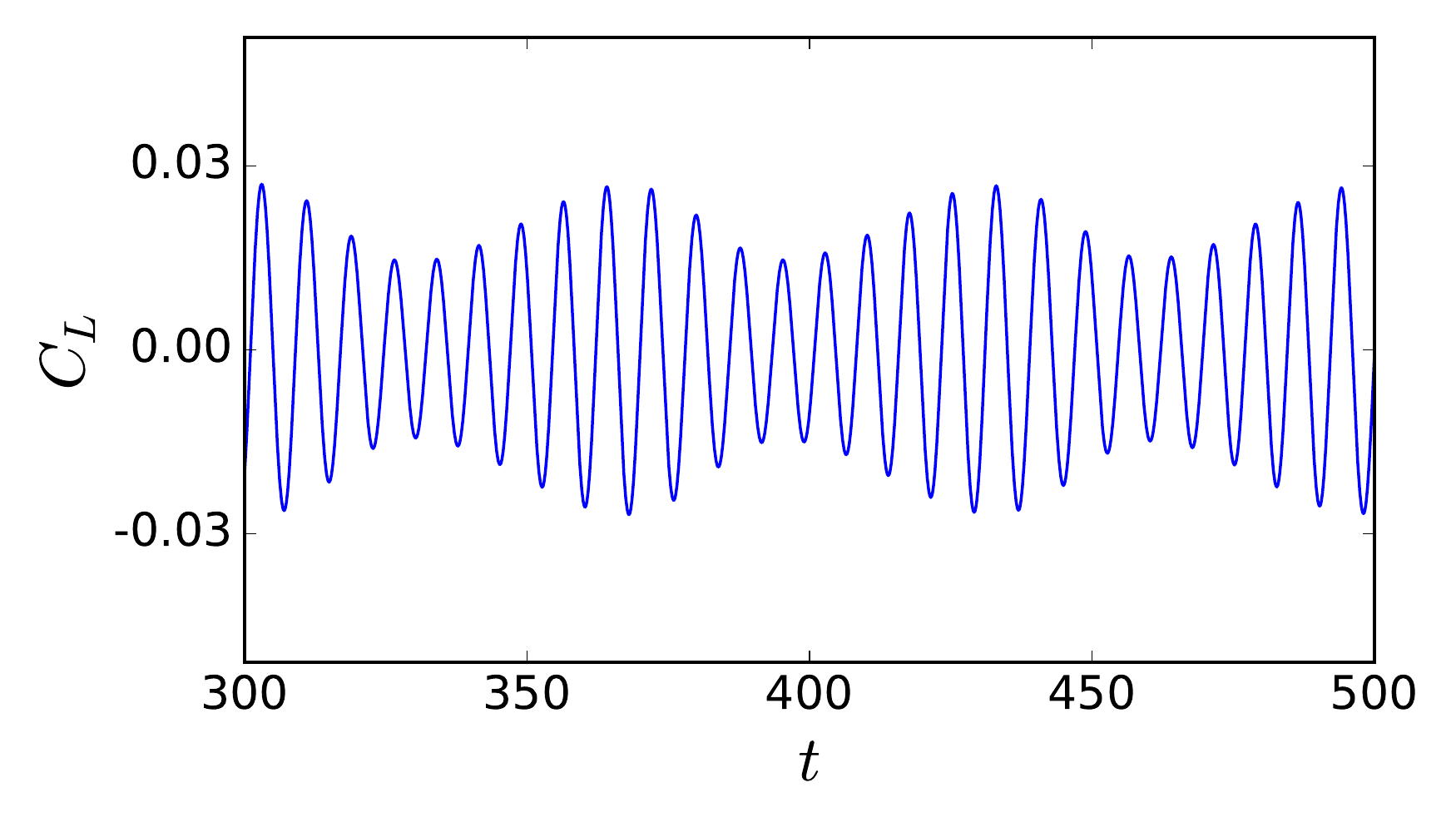}}
 \put(0.038,0.02){\includegraphics[width=.489\textwidth, trim={0 0 14.485 0},clip]{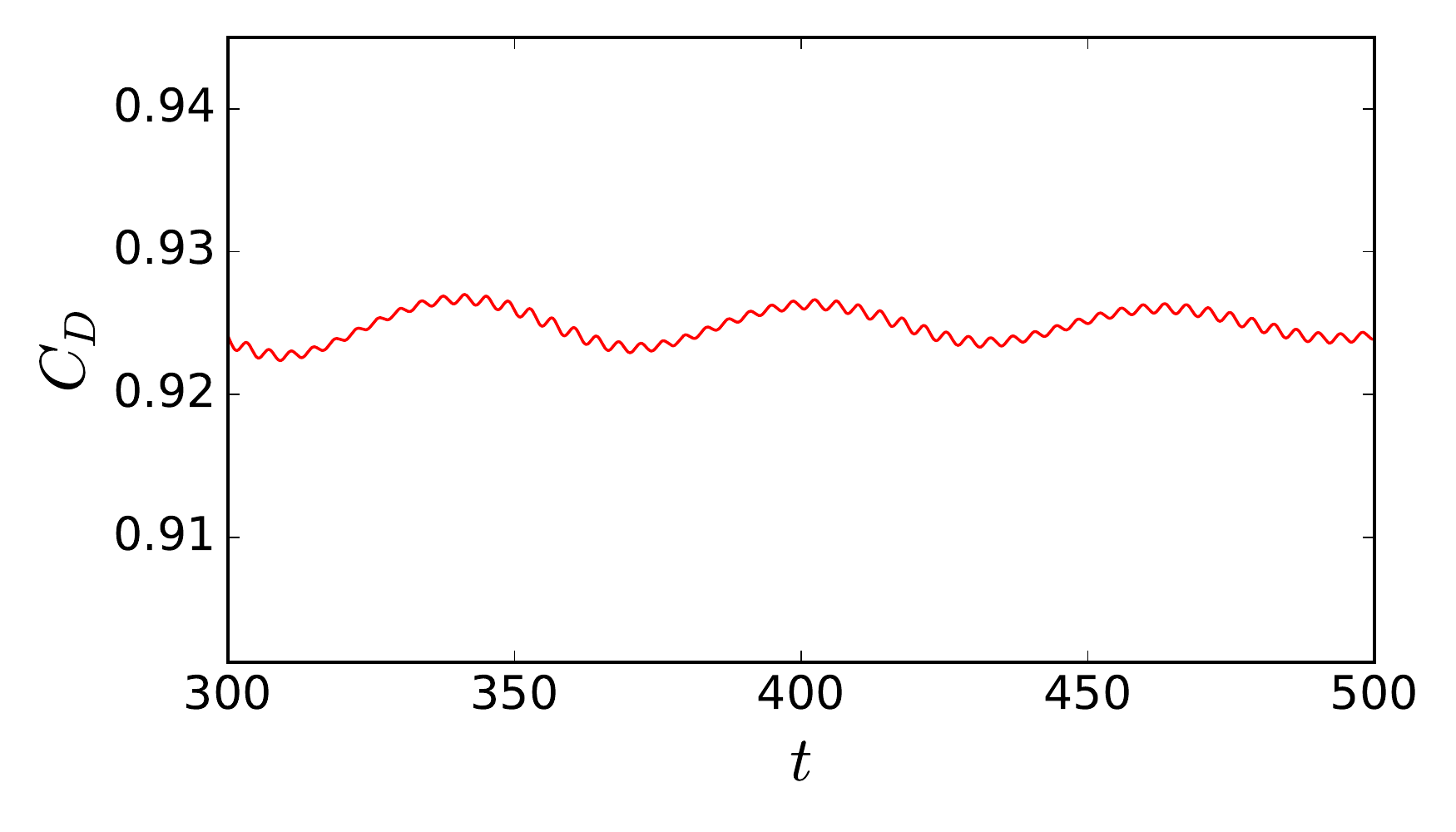}}
 \put(0.53,0.02){\includegraphics[width=.489\textwidth, trim={14 0 0 0},clip]{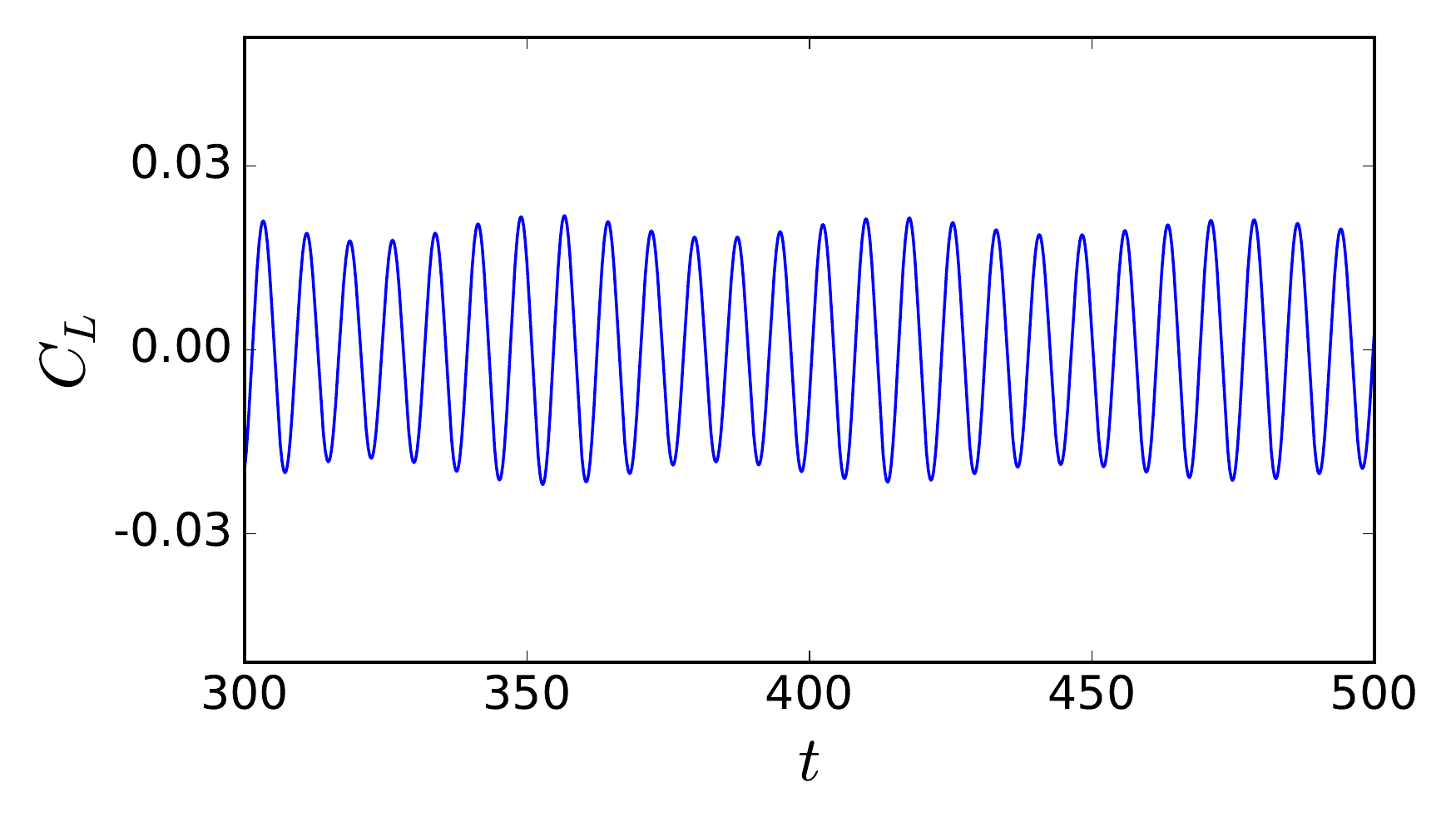}}
  \end{picture}
  \caption{\label{fig:timehistory-5.2}The time histories of $C_D$ and $C_L$ on the first cylinder for three-cylinder system with $p_1=5.2$ and a) $p_2=1.2$ b) $p_2=1.3$ c) $p_2=1.4$ and d) $p_2=1.5$. Note the drop in the mean drag and lift fluctuation between $p_2 = 1.2$ and $p_2 = 1.3$ as the lower bound is crossed - the range on the plots has been adjusted to show any variation beyond the lower bound.}
\end{figure}

\NH{The lower boundary is a reasonably ``hard'' transition, with the
  vortex shedding in the first gap remaining periodic when the third
  cylinder position is below the lower boundary, and an almost-steady
  flow occurs once the third cylinder moves beyond the lower
  boundary. The time histories of $C_L$ and $C_D$ for $p=5.2$ and
  $p_2=1.2, 1.3, 1.4$ and $1.5$ are presented in
  figure~\ref{fig:timehistory-5.2}. When $p_2=1.2$ - below the lower
  boundary - $C_L$ and $C_D$ are periodic. However increasing to
  $p_2=1.3$ sees a dramatic drop in the magnitude of fluctuation of
  the lift and drag coefficients. There is some modulation in the
  amplitude of $C_L$ and a variation in $C_D$ that occurs over a very
  long period which rapidly reduces further in magnitude as $p_2$ is
  increased further beyond the lower boundary.}

\begin{figure}
  \centering
\begin{tabular}{>{\centering\arraybackslash}m{2cm}m{2cm}m{2cm}}

 \hspace*{0cm}$\pthree=1.9$  &
 \hspace*{0.4cm}$\pthree=2.0$ &
 \hspace*{0.4cm}$\pthree=2.1$ \\

\hspace*{-0.2cm}\includegraphics[width=.17\textwidth, trim={20 50 260 50},clip]{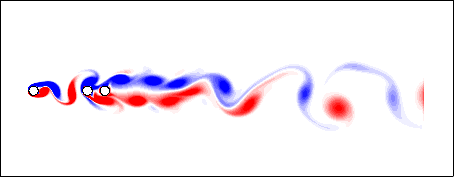}&
\hspace*{-0.2cm}\includegraphics[width=.17\textwidth, trim={20 50 260 50},clip]{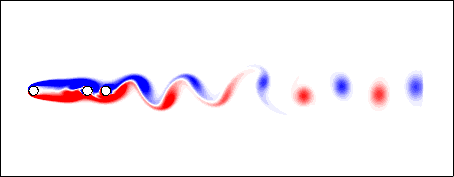}&
\hspace*{-0.2cm}\includegraphics[width=.17\textwidth, trim={20 50 260 50},clip]{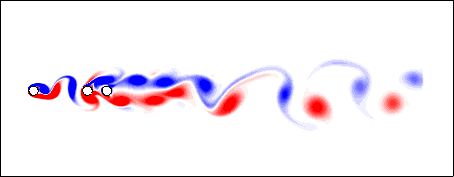} \\

\end{tabular}
 
\caption{\label{fig:FLow_vis_different_states58}Instantaneous flow
  visualization of the flow before, across and after the broadcasting
  region for $p=5.8$ and varying $\pthree=1.9, 2.0$ and $2.1$. All the
  visualizations are presented at the instant of the maximum lift on
  the second body.}
\end{figure}

The lower boundary is an increasing function of $p$. Figure
\ref{fig:FLow_vis_different_states58} shows a series of flow images
for $p = 5.8$. Here, with $\pthree = 1.9$, the flow is in the
convectively unstable structure; a slight increase to $\pthree = 2.0$
sees the appearance of the new global state which suppresses the
vortex shedding in the first gap; and a further slight increase to
$\pthree = 2.1$ sees the reappearance of the convectively unstable
structure.

This lower boundary is presented for $p \geq 5.2$ in
figure~\ref{fig:L_cPlot}. Despite the almost-constant upper boundary,
the lower bound dramatically increases such that broadcasting happens
only for a very short range of $\pthree$ when $p=5.8$, and the
feedback mechanism, and therefore the broadcasting, disappears for
$p\geq 6.0$. A perturbation in the region between the lower and upper
bounds leads to the broadcasting phenomenon.

\subsection{Characterisation of the new global mode}
\begin{figure}
  \centering
\begin{tabular}{>{\centering\arraybackslash}m{4.2cm}m{3.8cm}m{4.5cm}}

\hspace*{0.35cm} First cyl.&
\hspace*{0.9cm}Second cyl.&
\hspace*{1cm}Third cyl.\\

\hspace*{-0.3cm}\includegraphics[width=.345\textwidth, trim={15 0 64 0},clip]{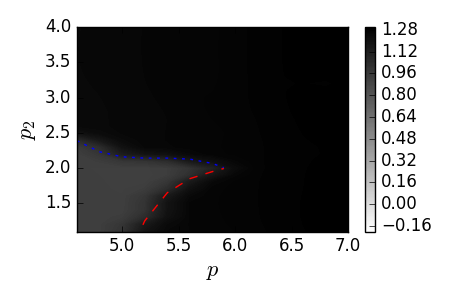}&
\hspace*{-0.07cm}\includegraphics[width=.29\textwidth, trim={53 0 64 0},clip]{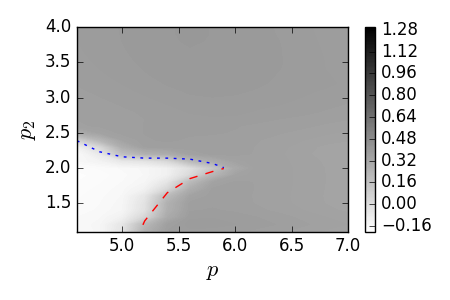}&
\hspace*{-0.13cm}\includegraphics[width=.363\textwidth, trim={55 0 9.99 0},clip]{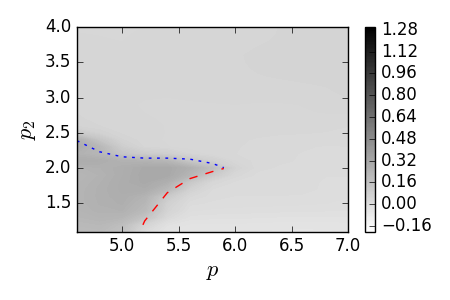}\\

\end{tabular}

\caption{\label{fig:Drag-contours}The contours of mean drag
  coefficient $\overline{C_D}$ as a function of $p$ and $\pthree$ for
  each cylinder. All the contours are produced using a common colour
  map. The dashed lines are the upper and lower boundaries obtained
  from figure~\ref{fig:L_cPlot}.}
\end{figure}

\begin{figure}
  \centering
\begin{tabular}{>{\centering\arraybackslash}m{4.2cm}m{3.8cm}m{4.5cm}}

\hspace*{0.35cm} First cyl.&
\hspace*{0.9cm}Second cyl.&
\hspace*{1cm}Third cyl.\\

\hspace*{-0.3cm}\includegraphics[width=.345\textwidth, trim={15 0 64 0},clip]{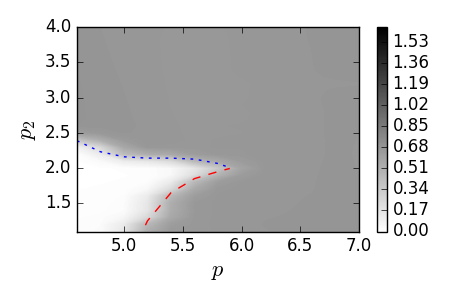}&
\hspace*{-0.07cm}\includegraphics[width=.29\textwidth, trim={53 0 64 0},clip]{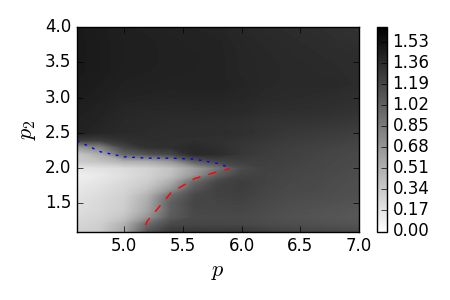}&
\hspace*{-0.13cm}\includegraphics[width=.363\textwidth, trim={55 0 9.99 0},clip]{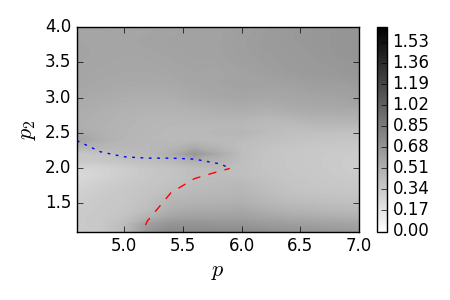}\\

\end{tabular}

\caption{\label{fig:lift-contours}The contours of maximum lift
  coefficient $C_{L_{\text{max}}}$ as a function of $p$ and $\pthree$
  for each cylinder. All the contours are produced using a common
  colour map. The dashed lines are the upper and lower boundaries
  obtained from figure~\ref{fig:L_cPlot}.}
\end{figure}

\begin{figure}
  \centering
\begin{tabular}{>{\centering\arraybackslash}m{4.2cm}m{3.8cm}m{4.5cm}}

\hspace*{0.35cm} First cyl.&
\hspace*{0.9cm}Second cyl.&
\hspace*{1cm}Third cyl.\\

\hspace*{-0.3cm}\includegraphics[width=.345\textwidth, trim={15 0 64 0},clip]{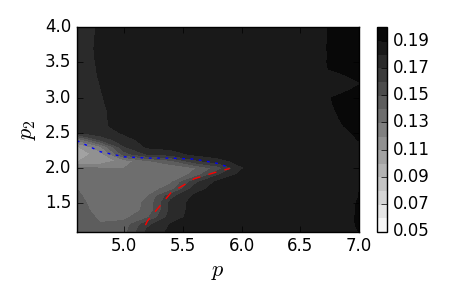}&
\hspace*{-0.07cm}\includegraphics[width=.29\textwidth, trim={53 0 64 0},clip]{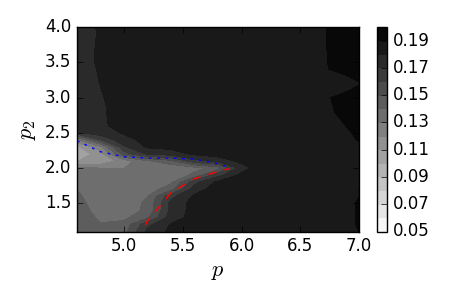}&
\hspace*{-0.13cm}\includegraphics[width=.363\textwidth, trim={55 0 9.99 0},clip]{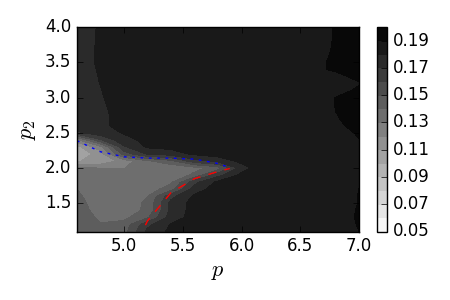}\\

\end{tabular}

\caption{\label{fig:Freq-contours}The contours of primary frequency
  $f$ as a function of $p$ and $\pthree$ for each cylinder.  All the
  contours are produced using a common colour map. The dashed lines
  are the upper and lower boundaries obtained from
  figure~\ref{fig:L_cPlot}.}
\end{figure}

The transition to a new global state and the broadcasting phenomenon
causes the flow to settle to a structure with no vortex shedding in
either the first or second gap, and a low frequency \karman\ wake
behind the third cylinder. This change in structure is also clearly
detected in the scalar measurements of the flow. Contour plots of mean
drag coefficient $\overline{C_D}$, maximum lift coefficient
$C_{L_{max}}$ and primary frequency $f$ with varying $p$ and $\pthree$
for each cylinder are presented in figures \ref{fig:Drag-contours},
\ref{fig:lift-contours} and \ref{fig:Freq-contours} respectively. The
acquired lines for the upper and lower boundaries of the broadcasting
region from figure~\ref{fig:L_cPlot} are also plotted on the
contours. These boundaries clearly coincide with the sudden change in
value of the contours of all the cylinders which is evidence of change
in structure of the flow and the forces.

Figure~\ref{fig:Drag-contours} shows that out of the broadcasting
region, $\overline{C_D}$ decreases from the first to the third
cylinder. However, in the broadcasting region there is a decrease in
the value of $\overline{C_D}$ from the first to the second cylinder
(in fact, at times the mean drag on the second cylinder is negative,
resulting in a slight thrust) while it increases again on the third
cylinder. Inside the broadcasting region the first and second cylinder
have a lower value of $\overline{C_D}$ than the third cylinder.

In contrast, all the cylinders experience a decrease in $C_{L_{max}}$
in the broadcasting region as shown in figure \ref{fig:lift-contours},
however this decrease is less significant for the third cylinder. This
is as expected for the observed flow structure - the lack of any
vortex shedding in the cylinder gaps means the fluctuating lift should
be small on the upstream cylinders, and the reduced wake width of the
modified \karman\ wake should also coincide with a reduced lift
force. Outside of the broadcasting region when the flow settles to the
two-row convective structure, the second cylinder has the higher
$C_{L_{max}}$ in comparison to the first and third cylinders. Again,
this fits with the flow structure - the second cylinder is exposed to
impinging vortices being shed from the first cylinder, and the
fluctuating wake.

The frequency contours in figure \ref{fig:Freq-contours} show that all
the cylinders have essentially similar values of primary frequency
either in the cloaking or broadcasting regions. In the cloaking
region, the frequency is set by the vortex shedding in the first gap,
and vortices at this frequency are simply convected past the second
and third cylinders. In the broadcasting region, a new global state is
triggered and all the three cylinders behave as almost one
more-streamlined body and therefore share a common frequency. This
common frequency is less than the frequency of vortex shedding from a
single cylinder or in the cloaking region.

\section{\NH{Sensitivity analysis of two-cylinder system}}

\begin{figure}
  \centering
\begin{tabular}{>{\centering\arraybackslash}m{1.4cm}m{6cm}m{6cm}}
&&\\
  &
\hspace*{0.6cm} Mean flow &
\hspace*{0.6cm} Sensitivity field\\
&&\\
\hspace*{-0.2cm} a) $p=4.8$&
\hspace*{-0.1cm}\includegraphics[width=.345\textwidth, trim={5 10 24 10},clip]{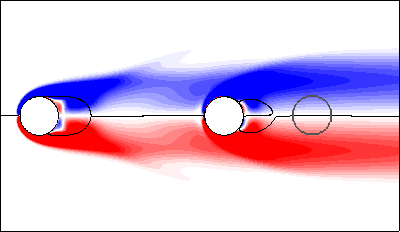}&
\hspace*{-1.6cm}\includegraphics[width=.535\textwidth, trim={5 10 24 10},clip]{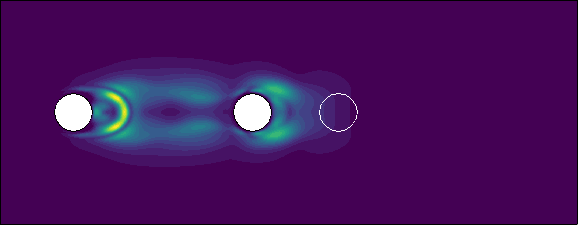}\\

\hspace*{-0.2cm} b) $p=5.2$&
\hspace*{-0.2cm} \includegraphics[width=.345\textwidth, trim={5 10 24 10},clip]{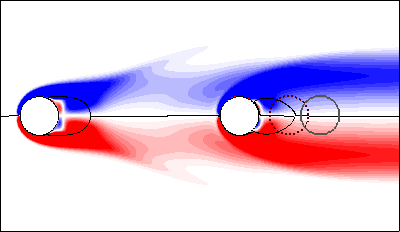}&
\hspace*{-1.6cm}\includegraphics[width=.535\textwidth, trim={5 10 24 10},clip]{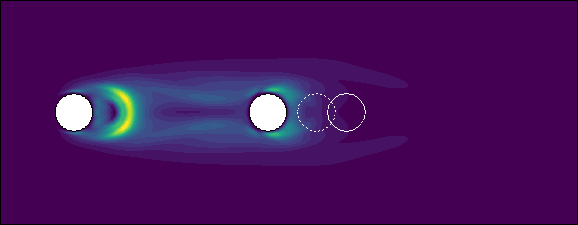}\\

\hspace*{-0.2cm} c) $p=5.6$&
\hspace*{-0.2cm} \includegraphics[width=.345\textwidth, trim={5 10 24 10},clip]{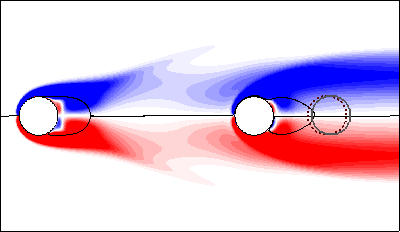}&
\hspace*{-1.6cm}\includegraphics[width=.535\textwidth, trim={5 10 24 10},clip]{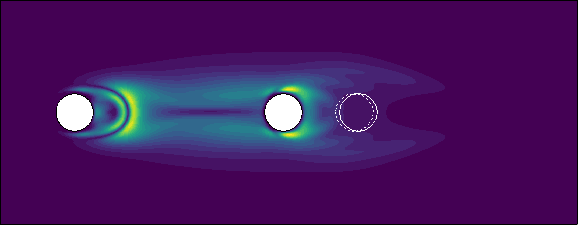}\\

\hspace*{-0.2cm} d) $p=6.0$&
\hspace*{-0.2cm} \includegraphics[width=.345\textwidth, trim={5 10 24 10},clip]{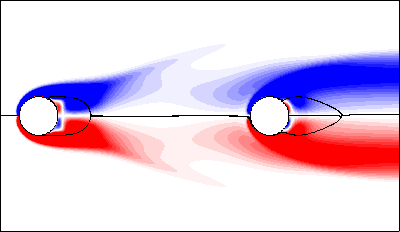}&
\hspace*{-1.6cm}\includegraphics[width=.535\textwidth, trim={5 10 24 10},clip]{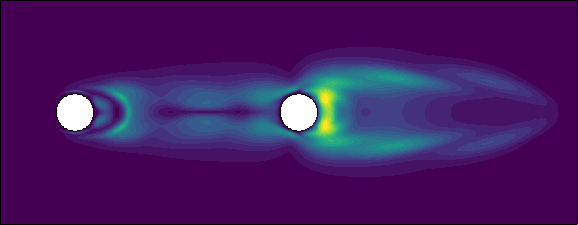}\\

\end{tabular}
\caption{\label{fig:sensitivity}Left: vorticity contours of the mean flow where red/blue contours mark positive/negative vorticity overlaid with separating streamlines. Right: contours of the sensitivity field for the leading global mode on this mean, where dark/light colours mark low/high values. Four cases for the two-cylinder system are shown with a) $p=4.8$, b) $p=5.2$, c) $p=5.6$, d) $p=6.0$. The dashed-line circles represent the location where the most upstream point of the circle coincides with the location of the lower boundry and the solid-line circles have their most upstream point coinciding with the location of the upper boundary as shown in figure~\ref{fig:L_cPlot}.}
\end{figure}

\NH{The sensitivity analysis outlined in section \ref{sec:sensitivity}
  is conducted on the mean flow of the two-cylinder system with
  different values of $p=4.8, 5.2, 5.6$ and $6.0$ in an effort to
  explain the presence of the lower limit for the third cylinder
  position to control the global mode. The mean flows for these values
  of $p$, and the sensitivity fields for the leading global mode on
  these mean fields, are presented in figure
  \ref{fig:sensitivity}. These values are chosen as they span three
  scenarios
  \begin{itemize}
  \item For $p = 4.8$, there is only an upper limit. Placing the third
    cylinder closer than this always suppresses vortex shedding in the
    first gap.
  \item For $p = 5.2$ and $5.6$, there is an upper and lower
    limit. Placing the third cylinder between these limits suppresses
    vortex shedding, but placing the third cylinder closer than the
    lower limit sees vortex shedding reinstated.
  \item For $p = 6.0$, there are no limits, and a third cylinder can
    be placed anywhere without any impact on the vortex shedding in
    the first gap.
  \end{itemize}
}

\NH{The first point to note from figure \ref{fig:sensitivity} is that the sensitivity field in the gap looks almost similar in all the cases - the change in the position of the second cylinder has little impact on the flow structure and therefore little impact on the mean and global mode. However, the structure of the sensitivity field behind the second cylinder varies for different $p$.}

\JL{Two main features are identified with increasing $p$. First, the location of the most sensitive region is initially located in two bands symmetric about the wake centreline, that slant from the most transverse points of the cylinder towards a point on the centreline around $2D$ downstream. These bands seem to track the location of the separating streamline. For $p = 4.8$, the most sensitive location is located around $1D$ downstream of the centre of the second cylinder, and around $0.6D$ transverse from the wake centreline. Increasing $p$ sees the most sensitive region move upstream and towards the centreline, so that at $p=5.6$ the most sensitive region is almost at the cylinder surface directly transverse of the cylinder centre. Increasing further to $p=6.0$ sees the most sensitive region move from these separating-streamline bands into the core of the mean recirculation region.}

\JL{The second feature is the development of a secondary sensitive region in the mean shear layers that grows in relative intensity with increasing $p$, however these regions never contain the most sensitive points in the flow. This shift in the location of the most sensitive region perhaps explains why no cylinder position can suppress or control the global mode when $p\geqslant 6$; once the sensitive regions move inside the mean recirculation region, any perturbation that triggers them has no mechanism to propagate to other regions of the flow as it is trapped inside the separating streamline. However, for lower $p$ where the sensitive regions are on or just outside the separating streamline, perturbations introduced on this streamline can be transported to the sensitive region, amplified, and transported back, forming a feedback loop that can lead to large growth and provide a physical mechanism to control the global mode and therefore broadcast the location of the third cylinder. This is consistent with the observed behaviour - the upper boundary of the third cylinder position that controls the global mode is almost constant at around $p_2 \simeq 2$, and the images in figure \ref{fig:sensitivity} show this coincides with a position where the most upstream point of the third cylinder comes into contact with a region of increased sensitivity aligned with the mean separating streamline.}

\JL{What is still not present is a clear indication of why the intermediate values of $p=5.2$ and $p=5.6$ present a lower limit for the location of a third cylinder that can suppress the global mode leading to the vortex shedding in the gap. There is some indication of the sensitivity field displaying a local maximum on the wake centreline which could coincide with the sensitivity (and therefore control) first rising, then falling again as the third cylinder is moved closer to the second. However, this local maximum is present in all the sensitivity fields whether or not a lower limit is identified, and so this condition does not seem sufficient to explain the presence of the lower limit.}

\JL{A more nuanced explanation may be provided by considering the fact that the current analysis is linear, and performed on the mean flow which itself is a function of the complete nonlinear flow. The third cylinder is certainly not a small perturbation, and will therefore induce nonlinear effects and therefore a mean flow correction. So, simply placing the third cylinder in regions where the two-cylinder mean flow is sensitive may not control the global mode, but instead remove the regions of sensitivity by modifying the mean flow in a way that these sensitive regions disappear.}
  
\begin{figure}
  \centering
\begin{tabular}{>{\centering\arraybackslash}m{6cm}m{6cm}}

a) & 
\hspace*{2.45cm} b) \\
\hspace*{0.059cm} \hspace*{-4.3cm}\includegraphics[width=.74\textwidth, trim={85 80 443 80},clip]{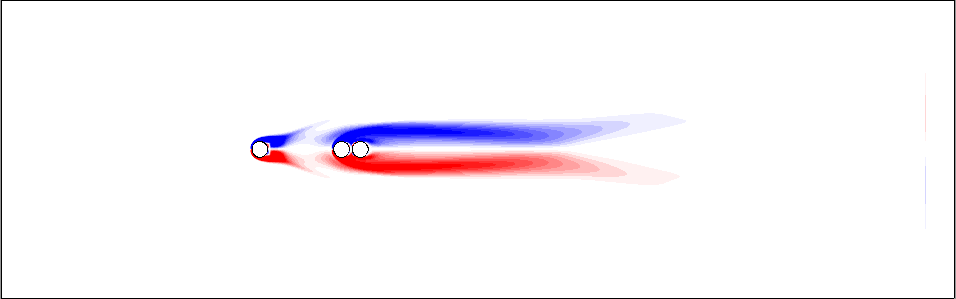}&

\hspace*{-0.24cm} \hspace*{-0.2cm}\includegraphics[width=.43\textwidth, trim={10 10 10 10},clip]{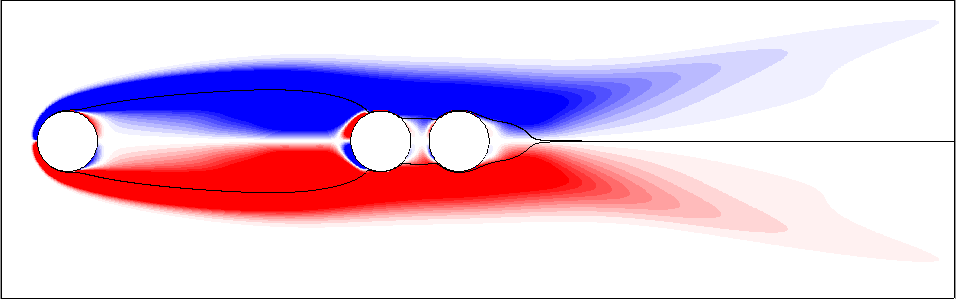}\\



\hspace*{-0.05cm} \hspace*{-0.7cm}\includegraphics[width=.46\textwidth, trim={10 10 10 10},clip]{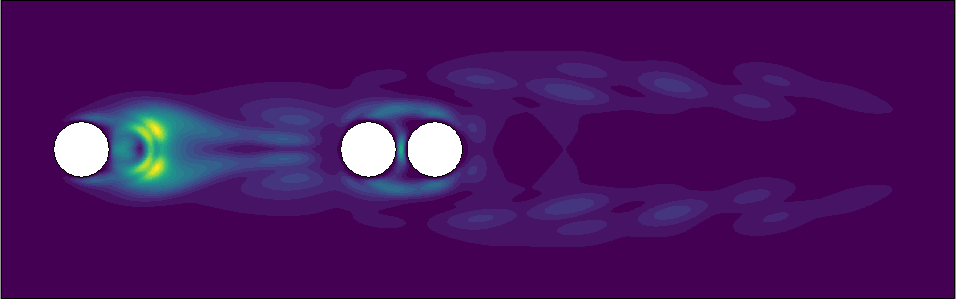}&
\hspace*{0cm} \hspace*{-0.39cm}\includegraphics[width=.52\textwidth, trim={10 28 10 28},clip]{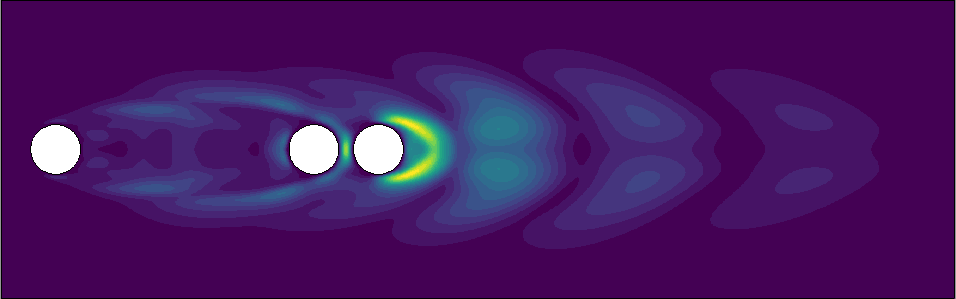}\\

\end{tabular}
\caption{\label{fig:mean-5.2}From top to bottom, the mean flow and sensitivity field for three-cylinder system with $p_1=5.2$ and a) $p_2=1.2$, b) $p_2=1.3$. }
\end{figure}

  
\NH{To further investigate this, the mean flow and sensitivity field for the three-cylinder system with $p_1=5.2$ and two critical locations for $p_2=1.2$ and $1.3$, before and after the lower boundary defined on figure \ref{fig:L_cPlot}, are studied and presented in figure~\ref{fig:mean-5.2}.}

\NH{As shown above in figure \ref{fig:FLow_vis_different_states52}, this minor change in location of the third cylinder has a drastic impact on the flow with periodic vortex shedding occuring in the first gap when $p_2 = 1.2$, and an almost-steady flow in this first gap when $p_2 = 1.3$. This global change is also reflected in the mean flow structure, and the subsequent sensitivity field which grows on this mean.}

\NH{The mean flow structure in the first gap for $p_2 = 1.2$ when vortex shedding occurs is very similar to that of the two-cylinder problem presented in figure \ref{fig:sensitivity}b, and accordingly the sensitivity field in this first gap is also similar. However, the presence of the third cylinder modifies the mean field such that the sensitive regions of the flow behind the second cylinder - which for the two-cylinder flow tracked the separating streamline down to the wake centreline - track the separating streamlines which enclose a region between the second and third cylinders. The effective streamlining of the two rearmost bodies acting as an elongated body breaks the feedback. }

\NH{For the case where $p_2 = 1.3$, the most sensitive region is clearly behind the third body, however there is also a highly sensitive region that forms between the second and third bodies and extends upstream along the separating streamlines. This is a consequence of the strong modification of the mean flow via the suppresion of the vortex shedding in the first gap.}

\NH{The observations provided in figure \ref{fig:mean-5.2} provide some indication that the impact of the third cylinder is not just to provide a disturbance in areas which are sensitive, but potentially to modify the mean flow such that these sensitive regions dissappear. This occurs when the third cylinder is very close to the second, forming an effective longer streamlined second body. Of course this conclusion comes with the caveat that the analysis performed is linear and the impact of the presence of the third cyinder has already modified the mean flow via a nonlinear correction.}



\section{Conclusions}
This study shows that a tandem two-cylinder system generates a two-row
shear layer structure once the pitch is long enough to allow vortices
to impinge on the second cylinder.

These shear layers simply convect the forcing imposed by the vortex
shedding from the first cylinder, maintaining its frequency and
spatiotemporal symmetry. This flow structure persists in larger arrays
of tandem bodies - bodies of varying size, shape and separation
distance can be included in the wake with little impact on the general
flow. Only by placing bodies close to the rear of the second cylinder
and perturbing an absolutely unstable region can this convective
structure be destroyed. Placing a third body near the very end of this
absolutely unstable region will perturb the absolutely unstable region
for $p \leq 5.8$. However, a lower limit also exists at least for
$p \geq 5.2$, meaning bodies placed too close to the rear of the
second body will not destroy the convective structure.

The destruction of the convectively unstable structure triggers a
global change in the flow structure that sees the three bodies behave
as a single, more streamlined body. No vortex shedding occurs in
either the first or second gap, and a modified, lower frequency
\karman\ wake occurs behind the third body. The transition to this new
global structure is clearly identified in the forces on all three
cylinders.
   
\section{Acknowledgement}

  NH acknowledges the support for Swinburne University of Technology
  via a Swinburne University Postgraduate Research Award (SUPRA). JL
  acknowledges the financial support of the Australian Research
  Council (ARC) via Discovery Project DP150103177, and the support of
  the National Computational Infrastructure, which is supported by the
  Australian government.

\bibliographystyle{jfm}
\bibliography{References}

\end{document}